\documentstyle{article}

\def\beq{\begin{eqnarray}}
\def\eeq{\end{eqnarray}}
\def\bea{\begin{eqnarray*}}
\def\eea{\end{eqnarray*}}

\def\centeron#1#2{{\setbox0=\hbox{#1}\setbox1=\hbox{#2}\ifdim
\wd1>\wd0\kern.5\wd1\kern-.5\wd0\fi
\copy0\kern-.5\wd0\kern-.5\wd1\copy1\ifdim\wd0>\wd1
\kern.5\wd0\kern-.5\wd1\fi}}
\def\ltap{\;\centeron{\raise.35ex\hbox{$<$}}{\lower.65ex\hbox{$\sim$}}\;}
\def\gtap{\;\centeron{\raise.35ex\hbox{$>$}}{\lower.65ex\hbox{$\sim$}}\;}

\def\eq#1{eq.~(\ref{#1})}

\newcommand{\newc}{\newcommand}
\newc{\qbar}{{\overline q}}
\newc{\Kahler}{K\"ahler }
\newc{\deltaGS}{\delta_{\rm GS}}

\begin{document}
\begin{titlepage}
\begin{flushright}
{\large hep-th/yymmnnn \\
SCIPP 10/07\\
}
\end{flushright}

\vskip 1.2cm

\begin{center}

{\LARGE\bf 
Supersymmetry and Its Dynamical Breaking}

\vskip 1.4cm

{\large Michael Dine$^a$ and John D. Mason$^b$}
\\
\vskip 0.4cm
{\it $^a$Santa Cruz Institute for Particle Physics and
\\ Department of Physics, University of California,
     Santa Cruz CA 95064  } \\

{\it $^b$Physics Department, Harvard University, Cambridge, MA  02138 } \\
\vskip 4pt

\begin{abstract}
This article reviews the subject of supersymmetry and its breaking. The emphasis is on recent developments
in metastable, dynamical supersymmetry breaking, which permit the construction of promising models of particle physics.
\vspace{1pc}
\end{abstract}

\end{center}

\end{titlepage}

\tableofcontents \clearpage

\section{Supersymmetry and Nature}

Supersymmetry has for some time been considered a candidate for physics beyond the Standard Model.
There are four reasons that are usually given that evidence for supersymmetry might appear at the TeV scale:
\begin{enumerate}
\item  Radiative corrections to the Higgs mass within the Standard Model are quadratically divergent.  If the Higgs mass is determined at some very high energy
scale, such as the Planck scale, a Higgs mass of order $M_Z$ requires that the ``bare" Higgs mass be extraordinarily fine-tuned.  In other words, the physical
mass is given by an expression of the form
\beq
m_H^2 = m_0^2 + \Lambda^2 \sum_{n=1}^{\infty} c_n \left ( {g^2\over 16 \pi^2 }\right )^n
\label{quadraticdivergence}
\eeq
(where $g$ denotes a generic coupling constant in the theory) and only by choosing $m_0^2$ to be of order the cutoff, $\Lambda^2$, and adjusting its value precisely to $30$ decimal
places or more, can the left hand side be of order the observed electroweak scale.
 In supersymmetric theories, provided the symmetry is unbroken, the quadratic divergences are absent, due to cancelations between Feynman diagrams involving
 bosons and fermions.  When supersymmetry is broken, $\Lambda$ in \eq{quadraticdivergence} is the scale of the breaking.  Avoiding fine tuning requires that this
 scale be not much more than a TeV (and arguably, as we will see, an order of magnitude less).
\item  With the assumption that all new thresholds lie at about $1$ TeV, the gauge couplings unify reasonably well, at a scale of order $10^{16}$ GeV.
\item  With two additional assumptions: a conserved $R$ parity, (the simplest hypothesis through which to forbid rapid proton decay), and that the gravitino is not the lightest
of the new supersymmetric particles, the theory
automatically possesses a dark matter candidate, produced in abundance comparable to the observed dark matter density.
\item  Supersymmetry arises rather naturally in many string constructions.
\end{enumerate}

But there is a fifth reason, first pointed out by E. Witten\cite{Witten:1981nf}:  supersymmetry would seem to
be poised to break dynamically.  In globally supersymmetric theories, supersymmetry is broken if the vacuum energy possesses {\it any} non-zero
value.  So even if supersymmetry is unbroken at tree level, one might expect all sorts of quantum effects to break
it.  However, quite generally, if there is a small, dimensionless coupling constant, $g$,  in the theory, supersymmetry is unbroken
to all orders in perturbation theory in $g$ if it is unbroken classically.  This is a consequence of a set of
{\it non-renormalization theorems}, which we will review.  These leave open the possibility that the
breaking can be of order $e^{-a/g^2}$, for some constant $a$.  This might account for exponentially large hierarchies.  That this phenomenon
indeed occurs will be a central focus of this review.

There are also good reasons to be skeptical that nature exhibits low
energy supersymmetry, most dramatically the absence of a Higgs boson below $115$ GeV or so (the so-called ``little hierarchy problem.")
It is likely that within two years, we will have
at least hints of supersymmetry from the LHC, or we will have confidence that there are no gluinos or squarks below about $700$ GeV (and possibly higher).  As we await these
developments, it seems worthwhile to review our understanding of supersymmetry, and perhaps more importantly, supersymmetry breaking.
The past few years have seen significant progress in this area.  As a result, from a {\it purely theoretical} perspective, the case for low energy supersymmetry
seems better than ever.

This article reviews our current understanding of dynamical supersymmetry breaking (DSB).  It is intended to be accessible to readers with only limited
experience with supersymmetry.  Section two provides a basic introduction to supersymmetry.  Section three explains the basic issues in supersymmetry
breaking at a conceptual level, presenting a supersymmetric version of quantum mechanics and the Witten index theorem.
Section 4 provides an overview of the dynamics of supersymmetric theories, especially supersymmetric QCD.
Section 5 discusses O'Raifeartaigh models, a class of theories which break supersymmetry at tree level; indeed, we will
think of these more generally as theories in which supersymmetry is linearly realized in the effective lagrangian, and spontaneously
broken at the level of the equations of motion.  Section 6 discusses the important phenomenological problem of mediating supersymmetry
breaking.   We distinguish two classes of mechanisms:  intermediate scale supersymmetry breaking (``gravity mediation") and gauge mediation, and discuss
recent developments in these areas.  Section 7 turns to dynamical supersymmetry breaking in four dimensions, considering models with stable and
metastable supersymmetry breaking.  A variety of approaches to model building are considered.  Section 9 concludes with speculations
as to the role of supersymmetry in nature.

\section{Basics of Supersymmetry}

There are a number of excellent books and review articles about supersymmetry; among them are \cite{Wess:1992cp,Gates:1983nr,Weinberg:2000cr,Baer:2006rs,Aitchison:2007fn,Binetruy:2006ad,Drees,Dine:2007zp}.
Here we will introduce enough about superfields and the superspace formulation to permit the construction of globally supersymmetric lagrangians.

\subsection{N=1 Supersymmetry}

In four dimensions, it is possible to have as many as
eight supersymmetries.  It is unlikely that theories with $N>1$
play any role in low energy physics.  First, such
theories are non-chiral.  Second, it is
virtually impossible to break supersymmetry in theories with $N>1$\footnote{In globally supersymmetry, it is easy
to prove that one cannot partially break higher $N$ supersymmetries to $N=1$; essentially the order parameter, which is the
vacuum energy, breaks {\it all} of the supersymmetries.  In most cases, one can readily
see that there is no potential which permits such complete breaking of the symmetry.  In supergravity theories, the
situation is more subtle\cite{Cecotti:1984rk,Louis:2009xd}, but this will not be germane to our concerns in this review.}.
The symmetries simply prevent one from
writing any term in the effective lagrangian
which could yield supersymmetry breaking.

So if nature is supersymmetric at scales comparable
to the weak scale, there is almost certainly
only one supersymmetry.  The basic supersymmetry
algebra is then\footnote{For superfields and spinors, we follow the notation of \cite{Wess:1992cp}.}
\beq
\{ Q_{\alpha}, \bar Q_{\dot \beta} \}
=2 \sigma^{\mu}_{\alpha \dot \beta} P_{\mu}.
\label{nequalsone}
\eeq
There is a straightforward
recipe for constructing theories with this
symmetry.
We will first consider the case of global supersymmetry;
later, we will consider the generalization to
local supersymmetry.   There are two irreducible
representations of the supersymmetry algebra containing
fields of spin less than or equal to one.  These
are the chiral and vector superfields.  Chiral fields
contain a Weyl spinor and a complex scalar; vector
fields contain a Weyl spinor and a (massless)
vector.

It is convenient to introduce an enlargement of space-time, known as {\it superspace}, to describe supersymmetric systems.
One does not have to attach an actual geometric interpretation to this space (though this may be possible) but can view it
as a simple way to realize the symmetry algebra of \eq{nequalsone}.  The space has four additional, anticommuting (Grassmann) coordinates,
$\theta_\alpha, \bar \theta_{\dot \alpha}$.    Fields (superfields) will be functions of $\theta, \bar \theta$ and $x^\mu$.  Acting on this
space of functions, the $Q$'s and $\bar Q$'s  can be represented as differential operators:
\beq
Q_\alpha = {\partial \over \partial \theta_\alpha} -i \sigma^\mu_{\alpha \dot \alpha}\bar \theta^{\dot \alpha} \partial_\mu;~~~~~
\bar Q^{\dot \alpha} = {\partial \over \partial \bar \theta_{\dot \alpha}} - i \theta^\alpha \sigma^\mu_{\alpha \dot \beta} \epsilon^{\dot \beta \dot \alpha} \partial_\mu.
\eeq
One can readily check that these obey the algebra of \eq{nequalsone}.  Infinitesimal supersymmetry transformations are
generated by
\beq
\delta \Phi =
\epsilon Q + \bar \epsilon \bar Q.
\eeq
It is also convenient to introduce a set of {\it covariant derivative}
operators which anticommute with the $Q_\alpha$'s, $\bar Q_{\dot \alpha}$'s:
\beq
D_\alpha = {\partial \over \partial \theta_\alpha} +i \sigma^\mu_{\alpha \dot \alpha}\bar \theta^{\dot \alpha} \partial_\mu;~~~~~
\bar D^{\dot \alpha} =- {\partial \over \partial \bar \theta_{\dot \alpha}} - i \theta^\alpha \sigma^\mu_{\alpha \dot \beta} \epsilon^{\dot \beta \dot \alpha} \partial_\mu.
\eeq

As mentioned
above, there are two irreducible representations of the algebra which are crucial to  understanding field theories with $N=1$ supersymmetry:
chiral fields, $\Phi$, which satisfy $\bar D_{\dot \alpha} \Phi = 0$, and vector fields, defined by the reality condition $V= V^\dagger$.  Both of these conditions are invariant under
supersymmetry transformations, the first because $\bar D$ anticommutes with all of the $Q$'s.
In superspace (using the conventions of \cite{Wess:1992cp}),
a chiral superfield may be written as
\beq \label{chiralsupfield}
\Phi(x,\theta)= A(x) + \sqrt{2}
\theta \psi(x) + \theta^2F + \dots
\label{chiralfield}
\eeq
Here $A$ is a complex scalar, $\psi$ a (Weyl) fermion,
and $F$ is an auxiliary field, and the dots denote terms containing derivatives.
More precisely, $\Phi$ can be taken to be a function of $\theta$ and
\beq
y^\mu = x^\mu -i \theta \sigma^\mu \bar \theta.
\eeq
Under a supersymmetry transformation with anticommuting
parameter $\zeta$, the component fields transform
as
\beq
\delta A= \sqrt{2} \zeta \psi,
\label{atransform}
\eeq
\beq
\delta \psi = \sqrt{2} \zeta F + \sqrt{2} i
\sigma^{\mu} \bar \zeta \partial_{\mu} A,~~~~~
\delta F= -\sqrt{2}i \partial_{\mu} \psi \sigma^{\mu} \bar
\zeta.
\label{psitransform}
\eeq
Vector fields can be written, in superspace, as
\beq
V= i \chi - i \chi^\dagger  +\theta \sigma^{\mu} \bar \lambda
A_{\mu} + i \theta^2 \bar \theta \bar \lambda
-i \bar \theta^2 \theta \lambda + {1 \over 2}
\theta^2 \bar \theta^2 D.
\eeq
Here $\chi$ is a chiral field.

In order to write
consistent theories of spin one fields, it is necessary to enlarge the
usual notion of gauge symmetry to a transformation of $V$ and the chiral
fields $\Phi$ by superfields.  In the case of a $U(1)$ symmetry, one has
\beq
\Phi_i \rightarrow e^{q_i \Lambda} \Phi_i
~~~~ V \rightarrow V - \Lambda- \Lambda^\dagger.
\label{abeliangauge}
\eeq
Here $\Lambda$ is a chiral field (so the transformed $\Phi_i$ is also chiral).
Note that this transformation is such as to keep
\beq
\Phi^{i \dagger} e^{q_i V} \Phi^i
\label{vtransform}
\eeq
invariant.  In the non-abelian case, the gauge transformation for $\Phi_i$ is as in
\eq{abeliangauge},
where $\Lambda$ is now a matrix valued field.  The transformation of $V$ is more complicated,
but is defined so that \eq{vtransform} remains valid, interpreting $V$ as a matrix valued field.

For the gauge fields, the physical content is most
transparent in a particular gauge (really a class
of gauges) know as Wess-Zumino gauge.  This gauge
is  analogous to the Coulomb gauge in QED.  In that case,
the gauge choice breaks manifest Lorentz invariance (Lorentz transformations
musts be accompanied by gauge transformations),
but Lorentz invariance is still a property of physical amplitudes.
Similarly, the choice of Wess-Zumino gauge
breaks supersymmetry, but physical
quantities obey the rules implied by the symmetry.  In this gauge, the
vector superfield may be written as
\beq
V=-\theta \sigma^{\mu} \bar \lambda
A_{\mu} + i \theta^2 \bar \theta \bar \lambda
-i \bar \theta^2 \theta \lambda + {1 \over 2}
\theta^2 \bar \theta^2 D.
\label{vectorfield}
\eeq
Here $A_{\mu}$ is the gauge field, $\lambda_{\alpha}$
is the gaugino, and $D$ is an auxiliary field.
The analog of the
gauge invariant field strength is a chiral field:
\beq
W_\alpha = -{1 \over 4}\bar D^2 D_\alpha V
\eeq
or, in terms of component fields:
\beq
W_{\alpha} = -i \lambda_{\alpha}
+ \theta_{\alpha}D -{i \over 2}
(\sigma^{\mu} \bar \sigma^{\nu} \theta)_{\alpha} F_{\mu \nu}
+ \theta^2 \sigma^{\mu}_{\alpha \dot \beta} \partial_{\mu}
\bar \lambda^{\dot \beta}.
\label{wdefinition}
\eeq
In the non-Abelian case, the fields $V$ are matrix valued, and transform under
gauge transformations as
\beq
V \rightarrow e^{-\Lambda^\dagger}V e^{\Lambda}
\eeq
Correspondingly, for a chiral field transforming as
\beq
\Phi \rightarrow e^{\Lambda} \Phi
\eeq
the quantity
\beq
\Phi^\dagger e^V \Phi
\eeq
is gauge invariant.
The generalization of $W_\alpha$ of the Abelian case is the matrix-valued field:
\beq
W_\alpha = -{1 \over 4} \bar D^2 e^{-V} D_\alpha e^V,
\eeq
which transforms, under gauge transformations, as
\beq
W_\alpha \rightarrow e^{-\Lambda} W_\alpha e^{\Lambda}.
\eeq

To construct an action with $N=1$ supersymmetry, it is convenient
to consider integrals in superspace.  The integration rules are simple:
\beq
\int d^2 \theta \theta^2 = \int d^2 \bar \theta \bar \theta^2 = 1;~~~\int d^4 \theta \bar \theta^2 \theta^2 = 1,
\eeq
all others vanishing.  Integrals $\int d^4 x d^4 \theta F(\theta,\bar \theta)$ are invariant, for general functions
$\theta$, since the action of the supersymmetry generators is either a derivative
in $\theta$ or a derivative in $x$.  Integrals over half of superspace of {\it chiral} fields
are invariant as well, since, for example,
\beq
\bar Q_{\dot \alpha} = \bar D_{\dot \alpha} +2 i \theta^\alpha \sigma^\mu_{\alpha \dot \alpha} \partial_\mu
\eeq
so, acting on a chiral field (or any function of chiral fields, which is necessarily chiral), one obtains a derivative in superspace.
In order to build a supersymmetric lagrangian, one starts with a set of chiral superfields, $\Phi^{i}$,
transforming in various representations of some gauge
group ${\cal G}$.   For each gauge generator, there
is a vector superfield, $V^a$.  The most general
renormalizable lagrangian, written in superspace, is
\beq
{\cal L} = \sum_i \int d^4 \theta
\Phi_i^{\dagger} e^V \Phi_i + \sum_a
{1 \over 4 g_a^2} \int d^2 \theta W_{\alpha}^2
+c.c. + \int d^2 \theta W(\Phi_i) + c.c.
\label{superspacel}
\eeq
Here $W(\Phi)$ is a holomorphic function of chiral superfields
known as the superpotential.

In terms of the component fields, the lagrangian
includes kinetic terms for the various fields (again in Wess-Zumino gauge):
\beq
{\cal L}_{kin} = \sum_i \left ( \vert
D \phi_i \vert^2 + i\psi_i \sigma^{\mu} D_{\mu}
\psi_i^*  \right )
-\sum_a{1 \over 4 g_a^2}\left ( F_{\mu \nu}^{a}F^{a\mu \nu} -i \lambda^a \sigma^{\mu}
D_{\mu} \lambda^{a*}  \right ).
\label{kineticterms}
\eeq
There are also Yukawa couplings of ``matter" fermions (fermions in chiral multiplets) and scalars, as well as Yukawa couplings of
matter and gauge fields:
\beq
{\cal L}_{yuk} =
 i \sqrt{2}\sum_{ia}
( g^a \psi^i T_{ij}^a \lambda^{a} \phi^{*j} + c.c.){
+ \sum_{ij} {1\over 2}{\partial^2 W \over \partial \phi^i
\partial \phi^j} \psi^i \psi^j.}
\eeq
We should note here that we will often use the same label for
a chiral superfield and its scalar component; this is common practice,
but we will try to modify the notation when it may be confusing.
The scalar potential is:
\beq
V= \vert F_i \vert^2 + {1 \over 2} (D^a)^2.
\eeq
The auxiliary fields $F_i$ and
$D_a$ are obtained by solving their equations of motion:
\beq
F_i^{\dagger}=-{\partial W \over\partial \phi_i}~~~~~
D^a = g^a \sum_i \phi_i^* T_{ij}^a \phi_j.
\label{fcomponent}
\eeq.

Since the work of Ken Wilson long ago, we have become accustomed to the idea that quantum field theories should be thought of as {\it effective} theories, appropriate to
the description of physics below some energy cutoff (or at distance scales large compared to some characteristic distance scale).   In this case, there is no
restriction of renormalizability.  On the other
hand, one often {\it does} want to expand the
lagrangian in powers of momenta.  It is a simple
matter to generalize eq.~(\ref{superspacel}) to include
arbitrary terms with up to two derivatives:
\beq
{\cal L} = \sum_i
\int d^4 \theta K(\Phi^{i*}, \Phi^i)
+ \sum_{ab}\int d^2 \theta
{\rm f_{ab}}(\Phi) W_{\alpha}^a W^{\alpha b} +c.c.
+ \int d^2 \theta W(\Phi_i) +c.c.
\label{nonrenormalizable}
\eeq
The functions $W$ and $f$ are holomorphic
functions of the chiral fields (otherwise the last
two terms are not supersymmetric); $K$
is unrestricted.  It is not
hard to generalize the component lagrangian.  Note that
among the couplings are now terms:
\beq
{1 \over 4} \rm Re~ f_{ab}(\phi_i)
F_{\mu \nu}^a F^{\mu \nu b} + {1 \over 4}
{\rm Im} ~f_{ab}(\phi_i)
F_{\mu \nu}^a \tilde F^{\mu \nu b} + {\partial f_{ab}
\over \partial \phi_i} F_i \lambda^a \lambda^b.
\label{sampleterms}
\eeq

\subsection{$R$ Symmetries}

In supersymmetry, a class of symmetries known as $R$-symmetries are of great interest.  Such symmetries can be continuous or discrete.  Their
defining property is that they transform the supercharges,
\beq
Q_\alpha \rightarrow e^{i \alpha} Q_\alpha;~~~~Q_\alpha^* \rightarrow e^{-i \alpha} Q_\alpha^*.
\eeq
Necessarily, the superpotential transforms as $W \rightarrow e^{2 i \alpha}W$ under any such symmetry.

The action of an $R$ symmetry is particularly simple in superspace, where it rotates the $\theta$'s and $\bar \theta$'s oppositely.  For a continuous
$R$ symmetry, $\theta \rightarrow e^{i\alpha} \theta$.    Correspondingly,
if the scalar component of a chiral superfield $\Phi$ transforms with charge $r$ under the symmetry, the fermion transforms with charge $r-1$, and the
auxiliary field with charge $r-2$.  Gauginos transform with charge $1$ and gauge bosons are neutral; the superfield $W_\alpha$ transforms with charge $1$,
while the superpotential transforms with charge $2$.

Discrete $R$ symmetries can be thought of as continuous $R$ symmetries, with particular values of the phase $\alpha$ (e.g. $\alpha = e^{2 \pi i \over N}$ for
a $Z_N$ symmetry).  It is generally believed that continuous global symmetries do not arise in theories which can be consistently coupled to
gravity\footnote{For  a recent discussion of this question, which reviews the earlier literature, see \cite{Banks:2010zn}}, but discrete symmetries (which can be, and probably necessarily are, discrete gauge symmetries) are more plausible.  For example, compactifying a ten-dimensional string theory
on a manifold ${\cal M}$ to four dimensions, a subgroup of the higher dimensional rotation group may be a symmetry of ${\cal M}$, and survive into the
low energy theory as a discrete symmetry.  Because rotations act differently on fermions and bosons, the symmetry is an $R$ symmetry (it rotates
the supercurrents).

Continuous $R$ symmetries are crucial to stable, spontaneous supersymmetry breaking, a result which follows from a theorem of Nelson and Seiberg.
which we will shortly describe \cite{Nelson:1993nf}.
If they arise in nature, such symmetries should be low energy accidents.
Discrete $R$ symmetries could give rise to such approximate symmetries.  They could also be important for generation of
mass scales and understanding the cosmological
constant, as we will see.  Discrete symmetries can themselves be gauge symmetries, in which case they are connected with the existence of cosmic strings\cite{Preskill:1991kd}.

\subsection{Supersymmetry Currents}

As for any continuous symmetry in quantum field theory, supersymmetry is associated with conserved currents; the supersymmetry charges are integrals of the time components of these
currents.
The Noether procedure can be used to derive conserved supersymmetry currents from the invariance of the action under infinitesimal supersymmetry transformations. However, as
is familiar from the case of the energy-momentum tensor, these conserved currents are not unique. They are defined up to an overall shift by ``improvement terms" which maintain conservation of the supersymmetry current. In the case of the energy momentum tensor, various criteria can be used
 to resolve the ambiguity\cite{Callan:1970ze}. As a simple example, consider a free theory of one chiral multiplet and canonical Kahler potential. In this case, applying the Noether procedure for the transformations in \eq{atransform} and \eq{psitransform} yields the following conserved currents
\beq
J^{\mu}_{\alpha} = \sqrt{2} (\sigma^{\nu} \bar{\sigma}^{\mu}\psi)_{\alpha}\partial_{\nu}A^* ~~{\rm and } ~~ J^{\mu \dot{\alpha}} = \sqrt{2} (\bar{\sigma}^{\nu} \sigma^{\mu}\bar{\psi})^{\dot{\alpha}}\partial_{\nu}A,
\eeq
where we have used the equations of motion. These currents are conserved: $\partial_{\mu}J^{\mu}_{\alpha} = \partial_{\mu} J^{\mu \dot{\alpha}} = 0$. But
certain ``improved" currents have more useful properties; in particular, they fit into supermultiplets.  The ``improved" currents \cite{Ferrara:1974pz}
\beq
J^{\mu~({\rm imp})}_{\alpha} = J^{\mu}_{\alpha} + \frac{\sqrt{2}}{3} (\sigma^{\mu}\bar{\sigma}^{\nu} - \sigma^{\nu}\bar{\sigma}^{\mu})_{\alpha}^{~\beta}\partial_{\nu}(A^*\psi_{\beta})
\eeq
and
\beq
J^{\mu{\dot{\alpha}}}_{({\rm imp})} = J^{\mu{\dot{\alpha}}} + \frac{\sqrt{2}}{3} (\bar{\sigma}^{\mu}\sigma^{\nu} - \bar{\sigma}^{\nu}\sigma^{\mu})^{\dot{\alpha}}_{~\dot{\beta}}\partial_{\nu}(A\bar{\psi}^{\dot{\beta}})
\eeq
are also conserved. Using the equations of motion we can write the improved supersymmetry current as
\beq \label{thetacomp}
J^{\mu~({\rm imp})}_{\alpha} = \sqrt{2} \left( \frac{1}{3}(\sigma^{\nu}\bar{\sigma}^{\mu}\psi)_{\alpha}\partial_{\nu}A^* + \frac{2}{3}[A^*\partial^{\mu}\psi_{\alpha}]_- \right)
\eeq
where $[x \partial^{\mu}y]_- = (\partial^{\mu}x)y - x(\partial^{\mu}y)$. A similar expression hold for $J^{\mu\dot{\alpha}}$ . It is convenient to package these supersymmetry currents into one supercurrent superfield, ${\mathcal J}_{\alpha\dot{\alpha}} =  {\sigma}^{\mu}_{\alpha \dot{\alpha}}{\mathcal J}_{\mu}$. Up to an overall normalization, one can identify $J^{\mu~({\rm imp})}_{\alpha}$ in \eq{thetacomp} as the $\theta$-component of the supercurrent

\beq
{\mathcal J}^{\mu} = i \bar{\sigma}^{\mu \dot{\alpha} \alpha} \left( 2i \sigma^{\nu}_{\alpha \dot{\alpha}}[\Phi^*\partial_{\nu}\Phi]_- + D_{\alpha}\Phi \bar{D}_{\dot{\alpha}}\Phi \right),
\eeq
where $\Phi$ was defined in \eq{chiralsupfield}. Using the identity: $2(D_{\alpha}\Phi \bar{D}_{\dot{\alpha}}\Phi^*) -[D_{\alpha},\bar{D}_{\dot{\alpha}}]\Phi^*\Phi = 2i\sigma^{\mu}_{\alpha \dot{\alpha}} [\Phi^*\partial_{\mu}\Phi]_-$, one can write a more covariant expression for the supercurrent as (again up to an overall normalization)
\beq
{\mathcal J}_{\alpha \dot{\alpha}} = 2(D_{\alpha}\Phi \bar{D}_{\dot{\alpha}}\Phi^*) - \frac{2}{3}[D_{\alpha},\bar{D}_{\dot{\alpha}}]\Phi^*\Phi.
\eeq
Then it is straightforward to verify that
\beq
\bar{D}^{\dot{\alpha}}{\mathcal J}_{\alpha \dot{\alpha}} = D_{\alpha}X ~~~{\rm with}~~~ X = -\frac{1}{3}\bar{D}^2 \Phi^*\Phi.
\eeq
For theories with a general superpotential $W$ and Kahler potential $K$, the supercurrent has a straightforward generalization
\beq \label{genmodel}
{\mathcal J}_{\alpha \dot{\alpha}} = 2g_{ij}(D_{\alpha}\Phi^i \bar{D}_{\dot{\alpha}}\Phi^{*j}) - \frac{2}{3}[D_{\alpha},\bar{D}_{\dot{\alpha}}]K.
\eeq
where $g_{ij}$ is the Kahler metric.
A Ferarra-Zumino multiplet is defined to be any multiplet satisfying the equations
\beq \label{FZmultiplet}
\bar{D}^{\dot{\alpha}}{\mathcal J}_{\alpha \dot{\alpha}} = D_{\alpha}X ~~~{\rm with}~~~ \bar{D}_{\dot{\alpha}}X = 0.
\eeq
For the supercurrent in \eq{genmodel}, we can satisfy  \eq{FZmultiplet} by choosing $X =4W -\frac{1}{3}\bar{D}^2K$ and we find that this broad class of models have Ferarra-Zumino multiplets. However, as stressed in \cite{Komargodski:2009pc,Dienes:2009td,Komargodski:2010rb}, they are not always gauge invariant or uniquely defined on the entire Kahler manifold of the theory\footnote{The Ferarra-Zumino multiplet on different patches of the Kahler manifold may be related by non-trivial Kahler transformations} and this leads to new constraints on effective lagrangians for supersymmetric theories. In addition to the improved supersymmetry current, this multiplet also contains the improved stress-energy tensor. If the theory is conformal, the bottom component of the multiplet is the current associated with a superconformal R-symmetry. Since typical supersymmetric
 theories are not superconformal, the bottom component is not automatically conserved like the others. This formulation of the supercurrent is useful for describing non-linear realizations of supersymmetry \cite{Komargodski:2009rz}.

Whether or not the theory is superconformal,
in the presence of a $U(1)_R$ symmetry, there can also exist a supercurrent multiplet ${\mathcal R}_{\alpha \dot{\alpha}}$, whose
lowest component is the conserved $R$ current.   This multiplet satisfies
\beq
\bar{D}^{\dot{\alpha}}{\mathcal R}_{\alpha \dot{\alpha}} = \chi_{\alpha}~~~ {\rm with} ~~~ \bar{D}_{\dot{\alpha}}\chi_{\alpha} = D^{\alpha}\chi_{\alpha} -  D_{\dot{\alpha}}\chi^{\dot{\alpha}} = 0.
\eeq
In this case the bottom component of ${\mathcal R}_{\alpha \dot{\alpha}}$ is automatically conserved, indicating that the global $U(1)_R$ symmetry exists. This multiplet clearly does not exist for theories without R-symmetries.

Recently \cite{Komargodski:2010rb} a third multiplet (the S-multiplet, ${\mathcal S}_{\alpha \dot{\alpha}}$), the linear combination of the Ferarra-Zumino multiplet and the R-multiplet was considered. The defining equations are
\beq
\bar{D}^{\dot{\alpha}}{\mathcal S}_{\alpha \dot{\alpha}} = D_{\alpha}X  + \chi_{\alpha}
\eeq
with $X$ and $\chi_{\alpha}$ constrained as before.

\subsection{Non-Renormalization Theorems}

It has long been known that supersymmetric theories, in perturbation theory, exhibit remarkable properties \cite{Grisaru:1979wc}.
Most strikingly, from detailed studies of Feynman graphs, it was shown early on that the superpotential of $N=1$ theories is not renormalized.
Seiberg \cite{Seiberg:1994bz}, in a program that has had far reaching implications, realized that these theorems could
be understood far more simply.  Moreover, Seiberg's proof indicates clarifies when non-perturbative
effects might be expected to violate the theorems.  His ingenious suggestion was to consider the couplings in the superpotential,
and the gauge couplings, as expectation values of chiral fields.  These fields must appear holomorphically
in the superpotential and gauge coupling functions, and this greatly restricts the coupling dependence of
these quantities.

To illustrate, consider a simple Wess-Zumino model:
\beq
W = {1 \over 2} m \phi^2 + {1 \over 3} \lambda \phi^3.
\eeq
For $\lambda =0$, this model possesses an R symmetry, under which $\phi$ has $R$ charge $1$.
So we can think of $\lambda$ as a chiral field with $R$ charge $-1$.  Since the
superpotential is holomorphic, the only allowed terms, polynomial in the $\phi$'s, have the form
\beq
\Delta W = \sum_n \lambda^n \phi^{n+2}.
\eeq
This is precisely the $\lambda$ dependence of tree diagrams with $n+3$ external legs; we have predicted that there
are no loop corrections to the superpotential.
Note that there is no corresponding argument for the Kahler potential, and it is easy to check that the Kahler potential is already
renormalized at one loop.  As a result, physical masses and couplings {\it are} corrected in this model.


Gauge theories exhibit similar non-renormalization theorems.  To understand this, it is necessary to first include the $CP$ violating parameter,
$\theta$,
\beq
{\cal L}_{\theta} = {\theta \over 16 \pi^2} F \tilde F
\eeq
where
\beq
\tilde F_{\mu \nu} = {1 \over 2} \epsilon_{\mu \nu \rho \sigma} F^{\rho \sigma}.
\eeq
Defining
\beq
\tau = i {\theta} + {8 \pi^2 \over g^2}
\eeq
the Lagrangian may be written, in superspace:
\beq
{\cal L}_{gauge} = {1 \over 16 \pi^2} \int d^2  \theta~ \tau {\rm Tr}~  W_\alpha^2 + c.c.
\eeq
Here, again, we think of $\tau$ as the expectation value of a chiral field.


In perturbation theory, physics is invariant under shifts of $\theta$, $\theta \rightarrow \theta + \alpha$, where
$\alpha$ is a continuous parameter.
So we immediately see, for example, that the superpotential is not renormalized.  The superpotential must be
a holomorphic function of $\tau$, and must respect the shift symmetry, so it must be independent of $\tau$.

The gauge coupling functions exhibit a similar, and somewhat paradoxical feature,
 Because they are holomorphic functions of $\tau$, they can at most consist of the leading $1/\tau$ term,
 and a constant.  In other words, they can only be renormalized at one loop.  On the other hand,
 from explicit calculations, one knows that the two loop beta functions of these theories are non-trivial.
 The resolution of this paradox is subtle, and we will refer the interested reader to the literature\cite{Novikov:1983mt,Dine:2007zp}.

We will see in section \ref{fourddynamics} that these non-renormalization theorems do not hold, in general, beyond perturbation theory.
This is key to the possibility that supersymmetry breaking is often exponentially small in supersymmetric theories.
Indeed, our arguments suggest a particular form, in many cases, for non-perturbative effects.
 Beyond perturbation theory, as is familiar from QCD, the continuous symmetry of the theory under shifts in (Re) $\tau$ is broken,
 leaving only a discrete, $2 \pi$ shift symmetry ($\alpha = 2 \pi n$).  From this, we can take an immediate lesson about the form of possible
 non-perturbative effects.  If they are to respect the periodicity, the low energy effective superpotential must depend on $\tau$
 as
 \beq
 \delta W = \sum_n e^{-n\tau} f_n(\Phi),
 \eeq
 where $\Phi$ are the various fields.  This structure can lead to exponential hierarchies for
 small $g^2$.  As we will see, in the presence of strong dynamics in the low energy theory,
 fractional powers of $e^{-\tau}$ can arise.

\section{Dynamical Supersymmetry Breaking:  A First Look}

In perturbation theory in the Standard Model, corrections to the Higgs mass are quadratically
divergent.   If nature is supersymmetric, these divergences cancel; if it is approximately supersymmetric,
with supersymmetry broken at the electroweak scale, these contributions are of order the weak scale (modulo
coupling constants).   In this sense, supersymmetry readily solves the hierarchy problem; this sort of cancelation
is sometimes referred to as ``technical naturalness".

But there is more to the hierarchy problem than the problem of divergences.  Why should enormous hierarchies exist at all?
Witten\cite{Witten:1981nf} pointed out that supersymmetric theories are potentially prone to the appearance of large hierarchies.
This is a result of two features of these theories:
\begin{enumerate}
\item  The vacuum energy is an order parameter for supersymmetry breaking.  The appearance of any non-zero value of the energy, no matter how small,
means that the symmetry is broken.
\item  The non-renormalization theorems, which insure that, if supersymmetry is unbroken at tree level, it  is unbroken
to all orders of perturbation theory, since no corrections to the superpotential can be generated.
But this is not guaranteed beyond perturbation theory, and the sorts of arguments we have reviewed for the theorems
actually suggest circumstances in which exponentially small effects might arise.
\end{enumerate}

\subsection{Supersymmetric Quantum Mechanics}

Witten provided a particularly simple realization of this idea in the context of a supersymmetric version of quantum mechanics\cite{Witten:1981nf}.
He considered a system with two supercharges, $Q_i$, $i=1,2$:
\beq
Q_1 = {1 \over 2} (\sigma_1 P + \sigma_2 W(x))~~~Q_2 = {1 \over 2} (\sigma_2 p - \sigma_1 W(x)).
\label{susyqm}
\eeq
It is easy to check that
\beq
\{Q_i,Q_j\} = \delta_{ij} H,
\eeq
where
\beq
H= {1 \over 2}\left  (p^2 + W^2 + \hbar \sigma_3{dW \over dx} \right ).
\label{qmhamiltonian}
\eeq
This system (which can be written in superspace\cite{vanHolten:1981en}), exhibits many of the features we have discussed
in four dimensional field theories.  Consider, for example, the non-renormalization theorems.  Classically, provided $W$ has a zero
somewhere, the system has a supersymmetric ground state.  When there is a sensible perturbation theory, this state
remains at zero energy, to all orders.  To illustrate the idea, suppose that one has
\beq
W = \omega x + {\cal O} (x^2)
\eeq
 near the origin.  Then the potential is approximately
 \beq
 V = {1 \over 2} \omega^2 x^2.
 \eeq
 The ground state energy then receives a (bosonic) contribution ${1 \over 2} \hbar \omega$.  But there is also a ``fermionic" contribution (i.e. a contribution
 multiplying the operator $\hbar \sigma_3$) in the Hamiltonian.  This is precisely
 \beq
 \Delta E = \pm \hbar \omega.
 \eeq
 So there remains a zero energy state.

 In this simple system, it is not hard to determine what happens beyond perturbation theory, {\it exactly}.  The condition that there be a supersymmetric
 ground state is simply:
 \beq
 Q_i \vert \psi \rangle = 0.
 \eeq
 Consider, for example, $Q_1$.  Multiplying by $\sigma_2$ gives the equation
 \beq
 \left ( -i {d \over dx} + i \sigma_3 W \right ) \psi =0,
 \eeq
 or
 \beq
 \psi =  e^{\pm \int_{-\infty}^x dx W \sigma_3} \psi_0,
\eeq
where $\psi_0$ is a constant spinor.  For functions, $W$, for which $\vert W \vert \rightarrow \infty$ as $x \rightarrow \infty$, this is {\it normalizable} (for
one choice of the spinor) only if $W$
is an {\it odd} function of $x$.  In practice, for even functions, one finds that the ground state energy is (for small coupling) exponentially close to zero.  For
example, for $W$ which is quadratic in $x$, the non-zero energy of the zero energy state of perturbation theory results from tunneling processes\cite{vanHolten:1981en}.

\subsection{The Witten Index Theorem}
\label{wittenindex}

The possibility that supersymmetry is broken non-perturbatively leads to the speculation that many theories, particularly strong interacting ones, might dynamically
break supersymmetry\cite{Dimopoulos:1981au,Dine:1981za,Witten:1981nf}.  But it turns out that there are strong
constraints, which were first outlined by Witten.  If supersymmetry is spontaneously broken
in a theory, the spectrum must include a massless fermion, analogous to the Goldstone boson of ordinary,
bosonic symmetry breaking, known as a Goldstino.
The simplest requirement for supersymmetry
breaking, then, is that the theory must possess a massless
fermion\cite{Witten:1981nf}, to play the role of the Goldstino.  Thus supersymmetric theories which possess a mass gap will not break supersymmetry.
A much more non-trivial set of constraints arise from the ``Witten Index"\cite{Witten:1982df}.  The basic idea is very simple.  Consider some supersymmetric
 field theory; place the system in a box (say with periodic
boundary conditions, which respect supersymmetry) so that the number of states is countable.  In a box, the supersymmetry generators (charges) are well-defined,
so states with {\it non-zero} energy come in Fermi-Bose pairs:
\beq
Q_\alpha \vert B \rangle = \sqrt{E} \vert F \rangle.
\eeq
Zero energy states, on the other hand, need not be paired; there might be a unique, bosonic zero energy vacuum, for example.  This simple observation leads to the
consideration of the index $\Delta$:
\beq
\Delta = {\rm Tr} (-1)^F e^{-\beta H}.
\eeq
The factor $e^{-\beta H}$ is included to make $\Delta$ mathematically well-defined.
The index receives no contribution from non-zero energy states, because of the Fermi-Bose pairing.  As a result, it is independent of $\beta$.
In fact, $\Delta$ is independent of {\it all} parameters of the theory.\footnote{This argument assumes that the theory not qualitatively change behavior as the parameter
varies; such behavior does occur, for example, in supersymmetric field
 theories as, say, some parameter is set to zero.}   To see this, consider how the spectrum might change
as parameters are changed.  Some non-zero energy state might come down to zero, but because of the pairing noted above, if it is a boson, say, it must be accompanied by
a fermion.
Similarly, a zero energy state might acquire non-zero energy, but again, only if accompanied by a state of the
opposite statistics.  So the index is invariant; it can be thought of (and in many interesting applications
is) topological in character, and finds applications in mathematics as well as physics.

Now what can we learn from $\Delta$?  If $\Delta$ is non-zero, we know that there is a zero energy state, and supersymmetry is  unbroken.  If $\Delta$ is zero, there may
or may not be a (Fermi-Bose paired) zero energy state; we do not know.

The index may be calculated in many cases\cite{Witten:1981nf}, including the supersymmetric quantum mechanics models we encountered above.  Most interesting
are gauge theories.  Consider, in particular, a pure $SU(N)$ gauge theory.  At infinite volume, this is a strongly coupled theory,
but because $\Delta$ is independent of parameters, it can be reliably calculated.
Witten gave a rigorous computation of the index, but
also an appealing, heuristic derivation, which we repeat here.

Consider the theory in a very small box, so that the coupling is very small.  Then non-zero momentum states have energy of order $1/L$.  Zero momentum states, on the other
hand, are lighter, and we can focus on these.  Let's specialize to $SU(2)$ and take the gauge $A^{0a} =0$.   In this gauge, states
must be invariant under time-independent gauge transformations.   The lowest energy configurations are expected
to be dominated by $\vec A^a$'s which commute, so take $\vec A$ in the $3$ direction..  Such constant $\vec A^a$ are {\it almost} gauge transformations.
But gauge transformations in the box must be periodic, so at most we can take $\vec A^3$, say, to lie in the interval
\beq
0 < A_i < {2 \pi \over L}
\eeq
(one can see this by considering periodic gauge transformations or possible Wilson lines).  Then the bosonic part of
 the system is a rotor, with energy eigenstates of order $g^2/L$, and lowest
energy $0$.

Now consider the fermions.  Again, we take these in the $3$ direction, so they are not affected by the gauge field.  We have, then, 2 fermion creation and two fermion annihilation
operators, and would thus seem to have two fermionic states and two bosonic states.  But in $A^0 = 0$ gauge, we need to require gauge invariance of the states
under an isospin rotation of $360^o$ about the $x$ axis.  This flips the sign of the fermion operators, and leaves only the two bosonic states.  This argument generalizes
to $SU(N)$.

The result applies not only to the pure gauge theory, but, by the independence of parameters, to the gauge theory with massive matter.  Again, this is consistent
with our results for supersymmetric QCD.  It does {\it not} apply to the theory with {\it massless} quarks, as the asymptotic behavior of the squark potential
changes as $m \rightarrow 0$.

\section{Supersymmetric Dynamics in Four Dimensions}
\label{fourddynamics}

Before discussing dynamical breaking of supersymmetry, it is important to understood the dynamics of broad classes of
supersymmetric field theories which {\it do not} break supersymmetry.  Indeed, these theories are interesting in their
own right.
Supersymmetric field theories exhibit features which are quite surprising from the perspective of more familiar, non-supersymmetric
four dimensional theories.   Many of these features trace to the very restrictive structure of supersymmetric effective lagrangians,
along with the non-renormalization theorems.  What has proven most striking is the ability, exploiting these properties, to make
exact statements about the properties of these theories, even, in many
cases, when the theories are strongly coupled.  Supersymmetric generalizations of QCD, defined as $SU(N)$ gauge
theories with $N_f$ fields, $Q$, in the fundamental representation, and $N_f$ in the anti-fundamental, have proven a particularly
fertile ground for such studies, and we will focus on these here.  Other groups are discussed in a number of reviews and textbooks\cite{Intriligator:1995au,Peskin:1997qi,Terning:2006bq,Dine:2007zp}.

\subsection{Pure Gauge Theory:  Gaugino Condensation}

Consider a supersymmetric theory with $SU(N)$ gauge group, and no chiral fields.  In terms of component fields, such a theory consists
of a set of gauge fields and a set of gauginos, Weyl fermions in the adjoint representation.  Classically, and in perturbation theory,
this theory has a chiral symmetry under which
\beq
\lambda \rightarrow e^{i \alpha}\lambda.
\eeq
It is well-known that the shift symmetry of perturbation theory is anomalous, and is broken beyond perturbation theory to a discrete subgroup (for an introduction, see \cite{Dine:2007zp}, chapter 5).
In an $SU(N)$ gauge theory, for example, instantons contribute to an expectation value for\cite{Shifman:1985fi}:\footnote{This is known not to be the
complete answer, which, as we will describe later, can be obtained by weak coupling methods.  The discrepancy is discussed in \cite{Hollowood:1999sk,Davies:1999uw}}.
\beq
\langle (\lambda \lambda)^N \rangle  \propto e^{-{8 \pi^2 \over g^2} + i \theta} = e^{-8 \pi^2 \tau}.
\eeq
Such an expectation value would leave over a discrete $Z_N$ symmetry, and
would be consistent with our experience with ordinary QCD.  The first questions to ask about this theory are
\begin{enumerate}
\item  Is supersymmetry spontaneously broken?
\item  Is the discrete $R$ symmetry spontaneously broken?
\end{enumerate}
Witten has shown that in this theory, in fact, there are $N$ supersymmetric ground states\cite{Witten:1982df},
as we reviewed in section \ref{wittenindex}.
Reasoning by analogy to ordinary QCD, he conjectured that there is a non-zero gaugino condensate in this theory:
\beq
\langle \lambda \lambda \rangle = c \Lambda^3.
\eeq
This condensate need not be associated with supersymmetry breaking; it is the lowest component of the chiral superfield $W_\alpha^2$, and
is thus invariant under supersymmetry transformations.
In a more symmetric fashion, we can write:
\beq
\langle W_\alpha^2 \rangle = \langle \lambda \lambda \rangle = \Lambda^3 e^{i \theta/N} \propto e^{-3 \tau/b_0}.
\eeq
We will see shortly that this result -- and in fact the precise value of the condensate --
can be derived, following Seiberg\cite{Seiberg:1994bz}, by simple arguments.

One can think of this as a constant superpotential (it is the coefficient of $\int d^2 \theta$ in the
effective theory at scales below $\Lambda$), so it represents a breakdown of the non-renormalization theorems of perturbation theory.
By itself, one might argue that this is not so interesting.  In global supersymmetry, physics is not sensitive to a constant $W$ (though in local supersymmetry,
if one started in flat space, as we will see in section \ref{supergravity},  one would now have a theory with unbroken supersymmetry in anti-De Sitter space).  But now couple
the gauge theory to a singlet, $S$, with no other couplings:
\beq
{\cal L} = (\tau + {S \over M} ) W_\alpha^2.
\eeq
Classically, the theory has a moduli space; the field, $S$, has no potential.  But quantum mechanically,
\beq
W_{eff}(S)  \propto e^{-{\tau \over 3 b_0}}e^{-{1 \over 3 b_0}{S \over M} }.
 \eeq
 So a classical moduli space ($S$), is lifted;
we have breakdown of the perturbative
non-renormalization theorems and dynamical supersymmetry breaking through non-perturbative effects.
The potential for $S$ vanishes as $S \rightarrow \infty$; for small $S$, the coupling is strong, and the theory cannot be
studied by these methods.  One might conjecture about the existence of a stationary point in the potential near the origin, but there is no small parameter which might account for
metastability.  We will consider variants of this structure which do lead to metastable states with broken supersymmetry in section \ref{retrofitting}.

\subsection{Supersymmetric QCD}

Let's step back and think more about supersymmetric dynamics.
We first consider {\it Supersymmetric QCD} with $N_f$ flavors, which we will define to be a supersymmetric theory with gauge group $SU(N)$
and $N_f$  quarks in the $N$ and $N_f$ in the $\bar N$ representations, $Q_f, \bar Q_{\bar f}$.  Consider, first, the theory with massless quarks.
The model has a global symmetry $SU(N_f)_L \times SU(N_f)_R \times U(1)_B \times U(1)_R$.  Here we are listing only symmetries
free of anomalies.  $Q$ and $\bar Q$ transform as\footnote{We find it convenient to choose the baryon numbers of $Q$ and $\bar Q$ to be $\pm 1$, rather than
$1/N$; with this convention, ``baryons" have baryon number $\pm N$.}
\beq
Q:~(N_f,1,{1}, {N_f - N \over N_f})
\eeq
$$~~~~~~~ \bar Q:~(1,N_f,-1, {N_f - N \over N_f}).
$$
Let's check the cancelation of anomalies.  We are concerned about
triangle diagrams with the symmetry current at one vertex, and $SU(N)$ gauge bosons
at the other two vertices.    For the $SU(N_f)$ symmetry, the absence of anomalies is automatic, resulting from the
tracelessness of the $SU(N_f)$ generators; for $U(1)_B$ it follows immediately from the opposite baryon numbers of $Q$ and
$\bar Q$.  For the $R$ symmetry, since the $R$ charge of the gluino is $+1$, the gluinos make a contribution
to the anomaly proportional to $N$ (the Casimir of the adjoint representation).   The $R$ charges of the (fermionic) quarks and antiquarks,
$\psi_Q$ and $\psi_{\bar Q}$ are ${N_f - N \over N_f} -1 = -{N \over N_f}$.
So, as the Casimir of the fundamental is $1/2$, and there are $2 N_f$ fields of the same $R$ charge contributing to the anomaly,
we obtain cancelation.

It is important to understand the structure of the massless theory.  Classically, there
is a large moduli space of supersymmetric (SUSY) vacua.  The potential arises simply from the
$D^2$ terms of the gauge fields; it is enough to ensure that these vanish.   Up to gauge and flavor transformations,
for $N_f < N$,

\beq
Q = \left ( \matrix{v_1 &0 & 0 &\dots & 0 & \dots  \cr 0 & v_2 & 0 & \dots & 0 & \dots \cr & & & & &\dots \cr 0 & 0 & 0 & \dots & v_{N_f} & \dots  } \right )
\label{qvevs}
\eeq
and $Q = \bar Q$.

To see this
it is helpful, first, to write the auxiliary $D$ fields as $SU(N)$ matrices (see, for example, \cite{Dine:2007zp}, chapter 13)
\beq
D^i_j = Q^{*i} Q_j - \bar Q^i \bar Q^*_j - {1 \over N} \delta^i_j (Q^{*k} Q_k - \bar Q^k \bar Q^*_k).
\label{dtermcondition}
\eeq
Then note that the vev of $Q$ can be brought to the form of \eq{qvevs} by flavor and color
transformations.   If we set $\bar Q= Q$, examining \eq{dtermcondition},
the vanishing of the $D$  as above is automatic.  The $D^i_j$ operators are unaffected by flavor transformations on the $\bar Q$'s.
That this exhausts the solutions is seen simply by counting degrees of freedom.  Of the original $2 N N_f$ chiral fields, $N^2 - (N-N_f)^2$ are ``eaten"
by the broken generators; $N_f^2$ are Goldstone multiplets, so all of the multiplets are accounted for.  Alternatively, one can construct $N_f^2$ gauge invariant
chiral fields,
\beq
M_{f,\bar f}=\bar Q_f Q_f
\eeq
corresponding precisely to the parameter counting above.

We can focus on the light degrees of freedom, constructing an effective lagrangian for the $M_{f \bar f}$ fields.
Perturbatively, there is no superpotential,
 and these directions remain flat.    More generally, we can ask whether there is {\it any} superpotential one can write consistent with the symmetries.
There is a unique answer:
\beq
W ={ \Lambda^{3 N + N_f \over N - N_f} \over {\rm det} (M_{f,\bar f})^{1 \over {N-N_f}}}.
\label{wads}
\eeq
This respects the $SU(N_f)$ and non-anomalous $U(1)_R$ symmetries (it carries $R$ charge $2$).
One can go one step further\cite{Seiberg:1994bp}.  Introducing the background field $\tau$, which determines
the gauge coupling, one can assign it a transformation law, $\tau \rightarrow \tau + i~ C~ \alpha$, where the constant $C$ is such that
the $\tau$ transformation cancels the anomaly in the $R$ symmetry under which $Q$, $\bar Q$ have R charge zero.   Recalling the dependence of $\Lambda$ on
$\tau$, it is easy to check that \eq{wads} respects this symmetry as well.  Note that the superpotential gives rise to a ``runaway" potential, which tends
to zero at infinity.

One piece of evidence supporting the presence of the superpotential of \eq{wads} is obtained by introducing small masses for the quarks.
These masses:
\begin{enumerate}
\item   break the continuous $R$ symmetry down to a discrete $Z_N$ subgroup
\item eliminate the runaway behavior
\item yield ``nice" supersymmetric vacua.
\end{enumerate}
To illustrate this last point, take,
for simplicity, all of the masses equal, $m_f = m$,
and look for solutions respecting a vector-like $SU(N_f)$ flavor symmetry, $v_i = v$ in \eq{qvevs}.  Then $v$ satisfies an equation:
\beq
m v^{{2N_f \over N - N_f} + 2} = \Lambda^{3N - N_f \over N-N_f}.
\eeq
with solution
\beq
v^2 \propto \Lambda^2 {\Lambda \over m}^{N-N_f \over N} e^{2 \pi i k \over N}.
\eeq
Note that this expectation value breaks the $Z_N$ symmetry, and also tends to $\infty$ as $m \rightarrow 0$.  This is as expected from the Witten
index (both the number of vacua, for fixed, non-zero $m$, and the fact that the index is ill-defined as the asymptotic structure of the potential changes, i.e. as $m \rightarrow 0$).

In the case $N_f = N-1$, the expectation values in \eq{qvevs} completely break the gauge symmetry.
For large $v$, the theory is weakly coupled, with coupling $g^2(v)$.  If a superpotential is generated, it must be possible
to compute it in a systematic, weak coupling approximation.  Indeed one can; the superpotential is generated
by instantons, and can be calculated in a completely straightforward semiclassical analysis\cite{Affleck:1983mk,Dine:2007zp}.
Starting with this result, one can use holomorphy arguments\cite{Seiberg:1994bp} to determine the superpotential in cases with smaller numbers of flavors,
including the case $N_f = 0$.  To illustrate the procedure, consider the case of $SU(2)$ with a single flavor.  Introducing a small quark
mass, as above, breaks the discrete $R$ symmetry.  But what is particularly important here is that it generates an expectation value for the
superpotential,
\beq
\langle W \rangle = \left (\Lambda^5 m \right )^{1/2}.
\eeq
By holomorphy, this expectation value is independent of $m$.  More precisely, the massless theory possesses a continuous $U(1)_R$ symmetry.
Under this symmetry, $m$ has charge $4$.  So necessarily, $W$, which is holomorphic, has this same form for all $m$.  Suppose that $m \gg \Lambda$.  Then the effective
low energy theory is a pure $SU(2)$ gauge theory.  $W$ is the coefficient of $\int d^2 \theta$ in the lagrangian; in the low energy theory, this is $\langle \lambda \lambda \rangle$.  $(\Lambda^5 m)^{1/2}$ is, in fact, $\Lambda_{LE}^3$, i.e. it is the cube of the $\Lambda$ parameter of the {\it low energy} theory.  So we have verified the existence of
the gluino condensate {\it in the strongly coupled theory}, and actually computed its value in terms of the microscopic parameters!

This same type of argument allows the computation of the coefficient of the superpotential for any $N_f < N$.

\subsection{Generalizing Gaugino Condensation}

$SU(N)$ gauge theory without matter has a $Z_N$ discrete symmetry, broken by
a gaugino condensate, $\langle \lambda \lambda \rangle$, an order parameter of dimension $3$.
This can be readily generalized\cite{Dine:2009swa} to include order parameters of dimension one.

As an example, consider supersymmetric QCD with $N$ colors and $N_f$ flavors, $N_f < N$, and with
$N_f^2$ gauge singlet chiral fields, $S_{f,f^\prime}$.  For the superpotential, take:
\beq
W = y S_{f f^\prime} \bar Q_f Q_f^\prime +\lambda {\rm Tr} S^3.
\eeq
To simplify the writing, we have assumed an $SU(N_f)$ flavor symmetry; this is not necessary to any of our considerations
here.  This theory
possesses a $Z_{2(3N-N_f)}$ $R$ symmetry.
Calling
\beq
\alpha = e^{2 \pi i \over 6N-2N_f}
\eeq
we can take the transformation laws of the various fields to be:
\beq
\lambda \rightarrow \alpha^{3/2} \lambda~~~~~S_{f,f^\prime}
\rightarrow \alpha S_{f,f^\prime}~~~~ (Q,\bar Q) \rightarrow \alpha(Q,\bar Q).
\eeq
The superpotential rotates by $\alpha^3$ under this transformation; the instanton amplitude is invariant, as
can be seen by noting that an instanton produces $2N$ gaugino
 zero modes, and $2 N_f$ fermionic ($Q$ and $\bar Q$) zero modes.
 This symmetry is spontaneously broken by $\langle S \rangle;~ \langle \bar Q Q \rangle; ~\langle W_\alpha^2 \rangle$ and $\langle W \rangle$.

The dynamics responsible for this breaking can be understood along the lines of our earlier analyses of supersymmetric QCD.  Suppose, for example, that
$\lambda \ll y$.  Then we might guess that $S$ will acquire a large vev, giving large masses to the quarks.  In this case, one can integrate
out the quarks, leaving a pure $SU(N)$ gauge theory, and the singlets $S_{f,f^\prime}$.  The singlet superpotential follows by noting that the scale, $\Lambda$,
of the low energy gauge theory depends on the masses of the quarks, which in turn depend on $S$.  So
\beq
W(S) = \lambda S^3 + \langle \lambda \lambda \rangle_S.
\eeq
\beq
\langle \lambda \lambda \rangle = \mu^3 e^{-3{8 \pi^2 \over b_{LE} g^2(\mu)}}
= \mu^3 e^{-3{8 \pi^2 \over g_{LE} g^2(M)} + 3{b_0 \over b_{LE}} \ln(\mu/M)};
\eeq
\beq
b_0 = 3N - N_f;~b_{LE} = 3N.
\eeq

So
\beq
\langle \lambda \lambda \rangle = M^{3N-N_f \over N} e^{-{8 \pi^2 \over N g^2(M)}} \mu^{N_f \over N}.
\eeq
In our case, $\mu = yS$, so the effective superpotential has the form
\beq
W(S) = \lambda S^3 + (yS)^{N_f/N} \Lambda^{3-N_f/N}.
\eeq
This has roots
\beq
S = \Lambda \left ({y^{N_f/N} \over \lambda} \right )^{N \over 3N - N_f}
\eeq
times a $Z_{3N-N_f}$ phase.

Note that this analysis is self-consistent; $S$ is indeed large for small $\lambda$.
The dynamics in other ranges of couplings has alternative descriptions, but the result that the
discrete symmetry is spontaneously broken, while supersymmetry is unbroken, always holds, provided
there are sufficiently many chiral singlets and suitable couplings among them and to the quark superfields.

\subsection{$N_f \ge N_c$}

For $N_f \ge N_c$, in addition to the mesons, there also exist (anti-)baryons,
\beq
B^{i_{N_c+1}...i_{N_f}} = \frac{1}{N_c!}\epsilon_{i_1...i_{N_f}}\epsilon_{\alpha_1...\alpha_{N_c}}\left(Q^{i_1\alpha_1}...Q^{i_{N_c}\alpha_{N_c}}\right); \\
\tilde{B}^{i_{N_c+1}...i_{N_f}} = \frac{1}{N_c!}\epsilon_{i_1...i_{N_f}}\epsilon_{\alpha_1...\alpha_{N_c}}\left(\tilde{Q}^{i_1\alpha_1}...\tilde{Q}^{i_{N_c}\alpha_{N_c}}\right)
\eeq
which paramaterize the classical moduli space of the theory. Furthermore, the superpotential in \eq{wads} blows up in the weak couplings limit (as $\Lambda \rightarrow 0$) and therefore cannot be dynamically generated. Without perturbative or non-perturbative corrections to the superpotential, this implies that $N_f \ge N_c$ theories possess a moduli space of vacua in the full quantum theory. In the regime $N_f < 3 N_c$, where the theories are asymptotically free, there is a range of interesting phenomena. The behaviors of these theories were elucidated by Seiberg\cite{Seiberg:1994bz}, and can be divided into four cases:  $N_f = N_c$, $ N_f = N_c +1$ $N_c +2 \le N_f \le  {3 \over 2} N_c$,
and ${3 \over 2} N_c < N_f < 3 N_c$. It is useful to consider the behavior of the theory when mass terms are added for the quarks. In this case, we can view
 the masses, again, as chiral fields transforming under the symmetries $SU(N)_L \times SU(N)_R \times
 U(1)_B \times U(1)_A \times U(1)_R$.  Then holomorphy and the symmetries determine that:
\beq
\langle M \rangle_i^j \sim \Lambda^{(3N_c -N_f)/N_c} ({\rm det}~m)^{1/N_c}\left(m^{-1}\right)^i_j
\label{modmass}
\eeq
When $m \rightarrow 0$, the meson vev must lie on the moduli space of the quantum theory (quantum moduli space).

\subsubsection{$N_f = N_c$:  The Quantum Modified Moduli Space}
\label{nfequalsnc}

In the case that $N_f = N_c$, the moduli space can be parameterized by $N_f^2$ mesons, a baryon $B$, and an anti-baryon $\tilde{B}$. With a total of $N_f^2 +2$ holomorphic degrees of freedom, these over-count the dimension of the moduli space by one. Classically, one can construct the set of fields satisfying the conditions for vanishing
$D$ terms, \eq{dtermcondition}, and check that $B\tilde{B} - {\rm det}(M) = 0$; that there is such a constraint is not
 surprising, since $M$ and $B$ ($\tilde{B}$) are made of the same quark fields.
 The solutions are of two types.  The first is similar to that of \eq{qvevs}, with $Q = \bar Q$ a diagonal, $N_c \times N_c$ matrix.
 The second is simply
 \beq
 Q = v ~I;~~~~\bar Q = 0
 \eeq
 (or $Q$, $\bar Q$ reversed).   For either class of solutions, the constraint is readily seen to be satisfied.
 Counting the gauge invariant fields and allowing for the constraint yields the correct dimension for the moduli space.

 While the absence of a superpotential for the holomorphic variables indicates that the moduli space persists quantum mechanically, its form can be inferred by analyzing the massive $N_f = N_c$ theory in the limit that $m \rightarrow 0$.  \eq{modmass} yields: ${\rm det}(M) = \Lambda^{2N_c}$ and $B = \tilde{B} = 0$. Since this is not
  true on the classical moduli space, the classical and quantum moduli spaces must be different. By also considering baryon perturbations, one can determine the quantum modified moduli
  space to be
\beq
B\tilde{B} - {\rm det}(M) = \Lambda^{2N_c}.
\eeq
For large values of the fields, this can be verified by an explicit instanton computation \cite{Beasley:2004ys}.
A lagrange multiplier can then be used to implement this constraint in the effective theory. Note that the singular point at the origin of the moduli space of the classical theory has been removed so that there are now no points on the moduli space where new massless degrees of freedom can arise other than those paramaterized by the constrained mesons and baryons. So a description of the quantum theory is provided by\cite{Seiberg:1994bz,Seiberg:1994pq}:
\beq
\label{supqmms}
W = X(B\tilde{B} - {\rm det}(M) - \Lambda^{2N_c}).
\eeq

\subsubsection{$N_f = N_c +1 $:  ``s-confining"}

 For $N_f = N_c + 1$, the classical moduli space is defined by $N_f^2$ Mesons and $N_f$ baryons and anti-baryons subject to constraints which restrict the dimension of the moduli space to $N_f^2$: $\tilde{B}_iM^i_j = M^i_jB^j = 0$, and
\beq
B^{i_{N_f}}\tilde{B}_{j_{N_f}} = \frac{1}{N_c!} \epsilon_{i_1,..,i_{N_f}}\epsilon_{j_1,..,j_{N_f}}(M^{i_1}_{j_1}..M^{i_{N_c}}_{j_{N_c}}).
\eeq
In the classical theory, the quarks and
gluons become massless at the origin, giving rise to the singularity
on the moduli space.
 Unlike the $N_f = N_c$ case, taking the $m\rightarrow 0$ limit of \eq{modmass} indicates that the origin {\it is} part of the quantum moduli space. Since the quantum theory is strongly coupled there, the only description of the theory one has is in terms of the massless excitations of the constrained moduli (the mesons and baryons). In this case, the natural interpretation of the singularity is that the new massless particles present in this description are the components of the meson and baryons which, away from
 the origin, were removed by the classical constraint equations. A highly non-trivial check of this interpretation is provided by the 't Hooft anomaly constraints.  At the origin,
 the flavor symmetries are unbroken; the anomalies in the flavor currents can be shown to be the same computed in terms of the quarks and leptons and
 in terms of the mesons and baryons.
The classical constraints are reproduced by including an interaction term between the mesons and baryons:
\beq \label{supsconf}
W = \frac{1}{\Lambda^{2N_f -3}}\left( \tilde{B}_iM^i_jB^j -{\rm det}(M)\right).
\eeq
So, at the origin the full theory is one with $N_f^2 + 2N_f$ massless fields and superpotential \eq{supsconf}.  This is in some sense the simplest picture of a ``dual" theory. The electric quarks confine into uncharged mesons and baryons, which provide a weakly coupled description of the theory. This phenomenon is more generically referred to as ``s-confinement" \cite{Csaki:1996zb}. Here the mesons and baryons are dual weakly coupled variables that describe the IR of the original theory in the region of strong coupling. What makes this simple relative to what we will find for the cases with larger numbers of flavors is that there is such a direct correspondence between the moduli space in the region
 of weak coupling,  and the variables which provide a description of the IR limit of this theory at the origin. For larger numbers of flavors we shall find that the connection between the two theories is not as direct, only the moduli spaces of the two theories match.

There is a final consistency check, which makes this picture all the more compelling.  If
one adds a mass term for the $(N_c+1)$-th quark, and takes the mass large, the dual  theory should reproduce the sigma model of the $N_f = N_c$ theory, with dynamical scale $\tilde \Lambda$ satisfying $\tilde{\Lambda}^{2N_c}= m\Lambda^{2N_c-1}= m \Lambda^{2N_f-3}$. This indeed is what happens.  Once again, we invoke holomorphy
to first consider the case of small quark mass.  The mesons containing the massive quark are now massive.  Call
\beq
M_{N_f,N_f} = \Lambda^{2N_f-1} X.
\eeq
This yields the superpotential of \eq{supqmms}.  Taking, now, $m \rightarrow \infty$, $X$ becomes non-dynamical, playing the role of a lagrange multiplier.

\subsubsection{$N_c + 2 \le N_f \le  {3 \over 2} N_c$:  Seiberg Duality}

For larger numbers of flavors, the classical moduli space is again paramaterized by the baryons and mesons plus classical constraints. The classical constraint equations are again satisfied when one takes $m \rightarrow 0$ in \eq{modmass} indicating that the origin is in the quantum moduli space. However, unlike the previous case, the 't Hooft anomaly matching conditions are not satisfied by the un-constrained mesons and baryons. The failure of the anomaly matching conditions at the origin indicates that the baryons and mesons alone cannot characterize the physics at the origin. A different set of massless particles is required. Seiberg's proposal for a set of variables and a superpotential which satisfies the 't Hooft anomaly matching and parameterize the correct moduli space away from the origin is an $SU(N_f-N_c)$ gauge theory with $N_f$ quarks and anti-quarks, $q$($\tilde{q}$), and a meson transforming as a bi-fundamental under the global $SU(N_f)^2$ flavor group
 plus the superpotential:

\beq
\label{supdual}
W = \tilde{q}_iM^{ij}q_j.
\eeq
 We can readily understand this proposal if
 we first study the ${3 \over 2} N_c < N_f < 3 N_c$ theories. What will be important later on is that this dual description is free in the IR and this lends itself to reliable calculation.

Before going onto larger numbers of flavors, we note that these theories also satisfy an important consistency condition.  Upon adding a mass to one quark flavor in the original electric theory, the dual of this theory flows to the dual of a theory with $N_f' = N_f -1$ flavors and $N_c'=N_c$ colors. In particular, adding a mass term in the $N_f = N_c + 2$ theory should reproduce the $N_f = N_c + 1$ theory. In this case the heavy flavor appears as a linear term in the meson ($W = mM^{N_f,N_f}$) in the dual theory. The equations of motion for this field forces the $N_f$-th (anti-)quark to get a vev $\langle \tilde{q}_{N_f}q_{N_f} \rangle \sim m$, higgsing the dual $SU(2)$ theory at a scale $(m\Lambda)^{1/2}$. The remaining quark/baryon interactions become the corresponding interaction in the dual of the $N_f = N_c + 1$ theory. The ${\rm det}(M)$ interaction arises from an instanton of the completely broken dual $SU(2)$ gauge group. The instanton possesses four gaugino zero modes
and $2N_f$ fermion zero modes. The interaction $\frac{{\rm det}(M)}{\Lambda^{N_c+1}}$ piece arises as follows.  The $SU(2)$ instanton possesses four gaugino zero modes and $N_f$ $q$ and $\bar q$ zero modes.
$2N_f -4$ of the $q,\bar q$ zero modes are tied together with the $M$ Yukawa coupling.  This leaves two factors of $M_{f,\bar f}$ to be
explained.  These can each be thought of as a fermion times a boson.   The fermions are provided by two of the remaining $q,\bar q$ zero modes;
 the other pair combine with the gauginos, producing the remaining two scalars.  The factor of $\Lambda^{-(N_c+1)}$ arises from the
 $SU(2)$ beta function; an additional factor of $1/(m \Lambda)$ is obtained from the instanton scale size integral, which is infrared finite due to the ($m$- and $\Lambda$-dependent)
scalar $vev$.
 Rewriting the scale $\Lambda$ of the $N_f=N_c +2$ theory in terms of the scale $\Lambda_L$ of the $N_f = N_c+1$ theory ($\Lambda_L^{2N_c-1}=m\Lambda^{2N_c -2}$), one finds that \eq{supdual} plus the instanton contribution reproduce \eq{supsconf}.

\subsubsection{${3 \over 2} N_c < N_f < 3 N_c$:  The Conformal Window}

When the number of flavors exceeds ${3 \over 2} N_c$ a remarkable picture emerges. Begin by considering  a limit with $N$, $N_f$ large, and  $\epsilon = 1-3\frac{N_f}{N_c} \ll 1$. In this case there is a non-trivial fixed point at $N_cg_*^2 = \frac{8\pi^2}{3}\epsilon$. In this {\it Banks-Zaks} limit\cite{Banks:1981nn}, the dimension of the operator $(\tilde{Q}Q)$ is $D(\tilde{Q}Q) = \gamma + 2 = \frac{3(N_f-N_c)}{N_c}$, where $\gamma$ is the operator's anomalous dimension. Assuming that there still exists a non-trivial fixed point away from the $\epsilon \ll 1$ limit, gives a picture of the physics at the origin of moduli space for all theories in the range ${3 \over 2} N_c < N_f < 3 N_c$. First, all theories in this window flow to their conformal fixed points in the IR. Second, once the theories are assumed to be conformal, there exists the possibility that the quarks and gluons are massless at the origin and that they provide a suitable description of the physics there.

  Third, there exists
 a dual set of variables and a superpotential, which provide a description of the theory at the origin. While neither theory is ever a free theory, when one is near a strongly coupled fixed point the other is near a weakly coupled one. The dual sets of variables and superpotential are the same as mentioned for the magnetic free theory. They are the bifundamental meson under $SU(N_f)^2$ and $N_f$ (anti-)fundamental (anti-)quarks transforming under an $SU(N_f-N_c)$ gauge group with the superpotential \eq{supdual}. Note that this dual theory is just a different SQCD theory in the range ${3 \over 2} N_c < N_f < 3 N_c$, but with the addition of a meson and its associated superpotential. Furthermore, it is the unique theory in the conformal window which has the same number of baryonic directions paramaterizing its moduli space as the original theory.

 This discussion provides further insight into where the dual variables in the ``magnetic free" window originate. For $N_f < {3 \over 2} N_c$, if the theory were to
 flow to a non-trivial fixed point, the dimension of the meson,  $D(\tilde{Q}Q) <1$ would violate the unitarity bound. So, this cannot be what happens to theories in this regime. Rather, the theory defined by the dual variables becomes IR free.

\section{Supersymmetry Breaking}

For the question of supersymmetry breaking, $R$ symmetries play a crucial role, embodied in a theorem of Nelson and Seiberg\cite{Nelson:1993nf}:
In order that a generic lagrangian (one with all terms allowed by symmetries, not making the restriction of
renormalizability) break supersymmetry, the theory
must possess an $R$ symmetry; if the $R$ symmetry is broken, for a generic $W$, supersymmetry
is broken.   This theorem is easily proven by examining the equations ${\partial W \over \partial \phi_i}=0$.
First, for a generic superpotential, with no $R$ symmetry, the equations
\beq
{\partial W \over \partial \phi_i} = 0, ~i =1 \dots n
\label{wequations}
\eeq
are $n$ holomorphic equations for $n$ unknowns.  In general, these have solutions.  Ordinary symmetries do not change the counting.  One can always
consider invariant products of fields instead of fundamental fields, and the generic equations will have solutions.

Now suppose that one has an $R$ symmetry.  First, it is easy to see that \eq{wequations}, for generic $W$
consistent with the symmetry, need no longer have solutions.  Because the superpotential
has $R$ charge two, if a set of fields, $X_i$, have $R$ charge two, while the others, $\phi_a$, are neutral, $X_i$ appear only linearly in the superpotential.
If there more $X$'s than $\phi$'s, eqns. \ref{wequations} constitute more equations than unknowns, and supersymmetry is broken.
To see that $R$ symmetry breaking implies supersymmetry
breaking, consider a superpotential which is a polynomial in chiral fields, $\Phi_i$, $i=1,\dots N$, of $R$ charge
$r_i$.  The superpotential must possess $R$ charge two.  Necessarily, there is at least one field with positive $R$ charge, which,
by assumption, can be taken to have non-zero expectation value; call this
$\phi_n$.  Define $n-1$ fields, neutral under the $R$ symmetry:
\beq
X_i = {\phi_i \over \phi_n^{r_i/r_n}}.
\eeq
The superpotential then can be written in the form
\beq
W = \phi_n^{2/r_n} f(X_i).
\eeq
If $\phi_n \ne 0$, corresponding to broken $R$ symmetry, the equations for a supersymmetric minimum are:
\beq
\partial_i f = f =0.
\eeq
These are $N$ equations for $N-1$ unknowns, so supersymmetry is broken.

The issues are illuminated by certain broad classes of models
In general, $W$ has $R$ charge $2$, if $Q_\alpha$ has charge one.
Consider a theory with fields, $X_i$, $i=1,\dots N$ with $R=2$, $\phi_a$, $a = 1,\dots M$, with $R$ charge $0$.
Then the superpotential has the form:
\beq
W = \sum_{i=1}^N X_i f_i(\phi_a).
\eeq
Suppose, first, that $N=M$.  The equations ${\partial W \over \partial \Phi_i}=0$ are solved if:
\beq
f_i = 0;~~~X_i =0.
\eeq
(R unbroken, $\langle W \rangle = 0$.)
The first set are $N$ holomorphic equations for $N$ unknowns, and generically have solutions.  Supersymmetry is unbroken; there is a discrete
set of supersymmetric ground states; there are generically no massless states in these vacua.
The $R$ symmetry is also unbroken, $\langle W \rangle = 0$.

Next suppose that $N < M$.  Then the equations $f_i=0$ involve more unknowns than equations; they generally have an $M-N$ dimensional space of solutions,
known as a moduli space.  In perturbation theory, as a consequence of non-renormalization theorems, this degeneracy is not lifted.  There are massless
particles associated with these moduli (it costs no energy to change the values of certain fields).

If $N>M$,  the equations $F_i=0$ in general do not have solutions; supersymmetry is broken.  These are the O'Raifeartaigh models.
Now the equations ${\partial W \over \partial \phi_i}=0$ do not completely determine the $X_i$'s, and classically, there are, again, moduli.
Quantum mechanically, however, this degeneracy is lifted.

\subsection{O'Raifeartaigh Models}

The considerations of the previous section are illustrated by the simplest O'Raifeartaigh model\cite{O'Raifeartaigh:1975pr}:
\beq
W =  X ( \lambda A^2 - f) + m Y A.
\label{simplestor}
\eeq
The model possesses an $R$ symmetry with
\beq
R_Z = R_Y = 2; R_A = 0; X(\theta) \rightarrow e^{2 i \alpha} X(e^{-i\alpha} \theta), ~{\rm etc.}.
\label{ror}
\eeq
(In order that the renormalizable terms be the most general consistent with symmetries, it is also necessary
to suppose a $Z_2$ symmetry under which $Y \rightarrow -Y, A \rightarrow -A$ forbids $Y A^2$).
In this model, SUSY is broken; the equations:
\beq
{\partial W \over \partial X} = {\partial W \over \partial Y} = 0
\eeq
are not compatible.

If  $\vert m\vert^2 > \vert f\vert$, the vacuum has $\langle A \rangle =0 = \langle Y \rangle$; $X$ undetermined.
For $\langle X \rangle =0$, the fermionic components of $A$
combine with those of $Y$ to form a Dirac fermion of mass $m$, while the scalar components of $A$
have mass-squared $m^2 \pm \lambda F_X$ (the scalar components of $Y$ are degenerate with the fermion).
More generally, it is straightforward to work out the spectrum as a function of $\langle X \rangle$.

Quantum effects generate a potential for $X$. At one loop, this is known as the Coleman-Weinberg potential; the
calculation is explained below (section \ref{cwpotential}).  One
 finds that the minimum of the potential lies at $\langle X \rangle = 0$.  $X$ is lighter than other fields (by
a loop factor).  Because the scale of the potential for the scalar components of $X$ is parameterically smaller than that of the other fields,
they are referred to as ``pseudomoduli".  The spinor component of $X$ is massless; it is the Goldstino of supersymmetry breaking.
\beq
\langle F_X \rangle =-f^*
\eeq
is the decay constant of the Goldstino.

Up to this point, we have not mentioned another possible origin of supersymmetry breaking:  the Fayet-Iliopoulos mechanism\cite{Fayet:1974jb}.
In the case of a $U(1)$ gauge theory, a term in the lagrangian:
\beq
{\cal L}_{FI} = \xi \int d^4 \theta V
\label{fi}
\eeq
is gauge invariant.  Such a term can lead to a non-zero expectation value for an auxiliary $D$ field, and break supersymmetry.
However, Fayet-Iliopoulos terms are unlikely to be important in dynamical breaking.  The reason can be understood
by considering a very simple non-renormalization theorem:  if not present at tree level, such a term can be generated only
at one loop.  This point was originally made in \cite{Witten:1981nf}, and proven by means of Feynman graphs in \cite{Fischler:1981zk}.  However, the theorem
can readily be understood along the lines of our earlier discussion of non-renormalization theorems.  Any dependence on the gauge
coupling must arise holomorphically in $\tau = i {\theta} + {8 \pi^2 \over g^2}$.  But any non-trivial $\tau$ dependence in \eq{fi} would violate gauge invariance.
Such a one loop contribution arises only if the trace of the $U(1)$ generator is non-vanishing, implying a gravitational anomaly.
Indeed, general arguments suggest that Fayet-Iliopoulos terms cannot appear in theories which can be consistently coupled to gravity\cite{Komargodski:2009pc}.

\subsection{Aside:  The Coleman-Weinberg Potential}
\label{cwpotential}

The
basic idea of the Coleman Weinberg calculations for the pseudomoduli
potentials is simple.   At tree level, $\langle X \rangle$ (or simply $X$) label physically inequivalent, but
degenerate states.  But since the spectrum depends on the pseudomodulus vev, $\langle X \rangle$,
quantum mechanically the energy is a function of $X$.  To determine the leading contribution to the energy,
one first calculates the masses of particles as functions of the pseudomodulus.   To determine the energy,
one proceeds as in one's first elementary field theory course\cite{Peskin:1995ev}.
There is a contribution of ${1 \over 2} \hbar \omega$ for each bosonic mode, and minus ${1 \over 2} \hbar \omega$ for each fermion
(due to filling the Fermi sea).  As a result:
\beq
V(X) =\sum (-1)^F \int {d^3 k \over (2 \pi)^3}{1 \over 2} \sqrt{k^2 + m_i^2}
\label{vacuumenergy}
\eeq
where the sum is over helicity states, and $(-1)^F$ is $+1$ for bosons, $-1$ for fermions.
The separate terms are very divergent in the ultraviolet; expanding the integrand in powers of $k$ yields first a quartic divergence,
proportional to $\sum (-1)^F \Lambda^4$, where $\Lambda$ represents some cutoff; this vanishes, since the number of fermions and bosons are equal.  Next, there is a quadratic
divergence, proportional to $\sum (-1)^F m_i^2 \Lambda^2$.  This vanishes, as well, due to a sum rule, which holds at tree level:
\beq
\sum (-1)^F m_i^2 = 0.
\eeq
This sum rule holds in any globally supersymmetric theory with quadratic Kahler potential; it is proven, for example, in \cite{Dine:2007zp}.   Finally, there is a logarithmically divergent term.
This divergence is real; it is associated with the renormalization of the leading $F_X^\dagger F_X$ term in the potential, corresponding to a renormalization
of the kinetic term, $\int d^4 \theta X^\dagger X$.
It's value (and $X$ dependence) depends
on the details of the model.

Having understood the divergence structure, it is a simple matter to write a more precise formula.  Because the integral in \eq{vacuumenergy} is only logarithmically
divergent, it
can be evaluated using, for example, standard formulas from dimensional regularization\cite{Peskin:1995ev}:
\beq
V(X) = {1 \over 64 \pi^2} \sum (-1)^F  m_i^4 \ln(m_i^2).
\eeq
At small $X$, one finds that the potential grows quadratically with $X$; at large $X$, it grows logarithmically.   Asymptotically,
the potential grows logarithmically with $X$.  The $R$ symmetry is unbroken.
Shih has shown\cite{Shih:2007av}, quite generally, that if all fields in an O'Raifeartaigh model
have $R$ charge $0$ or $2$, then the $R$ symmetry is unbroken.  Shih constructed models
for which this is not the case.  One of the simplest theories which, for a range of parameters, breaks the $R$ symmetry spontaneously is:
\beq
W = X_2 (\phi_1 \phi_{-1} - \mu^2) + m_1 \phi_1 \phi_1 + m_2 \phi_3 \phi_{-1}.
\label{shihmodel}
\eeq

When the Goldstino decay constant, $f$, is much less than the characteristic mass scale of the theory ($f \ll m^2$ in our simple O'Raifeartaigh model, \eq{simplestor}), the analysis can be simplified, and framed in a language which exploits the approximate
supersymmetry of the system.  Integrating out physics at the scale $m$ leaves an effective action for $X$.  At scales large compared to $f$, this
action must be supersymmetric; supersymmetry breaking arises spontaneously within this effective action.  A simple argument for this
goes as follows.  Couple the system to gravity, then for energy scales large compared to $f$, the gravitino
has negligible mass, so consistency requires that the low energy theory must exhibit local supersymmetry (more discussion of supergravity
appears in section \ref{supergravity} below).

More directly, the action of such a model may be calculated using supergraph techniques\cite{Wess:1992cp,Grisaru:1979wc,Dine:1981za,Grisaru:1979wc}.
In the limit of small $f$, one can expand the effective action in powers of $f$.  A term of order $f^n$ in the
effective action can be obtained by computing diagrams with $n$ external $X$ legs.  The result has the form, for small $X$:
\beq
\delta {\cal L} = \int d^4 \theta {\lambda^2 \over 16 \pi^2}{1 \over M^2} X^\dagger X X^\dagger X.
\eeq
Substituting $X = X + \theta^2 f$ yields a quadratic potential for $X$:
\beq
V(X,X^\dagger) = {\lambda^2 \over 4 \pi^2} {\vert f \vert^2 \over M^2} \vert X \vert^2.
\eeq
More generally, the structure of the potential for $X$ can be understood in this fashion.  Evaluation of the relevant Feynman diagrams is, in fact,
quite simple.

\section{Mediating Supersymmetry Breaking}

The MSSM, by itself, does not break supersymmetry\cite{Affleck:1984xz}.  So, in attempting to
construct supersymmetric models of nature, it is necessary to include additional dynamics in order that
supersymmetry be broken.  The simplest possibility is to simply add an O'Raifeartaigh model, of the type we have discussed above.  It is then
necessary that some interactions transmit the breaking of supersymmetry to the ordinary fields.  The supersymmetry breaking sector of the
model is generally referred to as the ``hidden sector", while the ordinary fields of the MSSM are referred to as the ``visible sector".
Most model building for supersymmetry phenomenology assumes that supersymmetry breaking is transmitted, or mediated, through
gravitational strength interactions, and is thus referred to as ``gravity mediation".  In this case, as we will shortly explain, in order that
the effective scale of supersymmetry breaking be of order $1$ TeV, it is necessary
that the scale of supersymmetry breaking be of order
\beq
M_{int}^2 = M_p \times {\rm TeV} \approx (10^{11} ~{\rm GeV})^2.
\eeq
Models of this sort are said to be ``gravity mediated".  Models with lower scale are also of interest.  Successful models of this sort typically
involve gauge particles as the messengers, and are said to be ``gauge mediated."  In both cases it is necessary to understand some aspects
of the coupling of supersymmetric theories to general relativity, {\it Supergravity}.

\subsection{Supergravity}
\label{supergravity}

If a theory containing a graviton is supersymmetric, it is necessary that the graviton itself lies
in a supersymmetry multiplet.  The fermionic partner of the spin two graviton is a particle of spin $3/2$, the gravitino.
Just like a theory of a massless vector boson requires a gauge symmetry and a massless spin two field requires general covariance,
a massless spin 3/2 particle requires {\it supergravity}, a theory of gauged supersymmetry.  The construction of these theories
is rather complicated, and is described in a number of excellent textbooks\cite{Wess:1992cp,Weinberg:2000cr}.  We will only review a few
features here which will be important in our subsequent discussions.

Much like the global case, the general supergravity
lagrangian is specified by a Kahler potential, $K(\phi_i,\phi_i^\dagger)$, superpotential, $W(\phi_i)$,  and gauge coupling functions, $f_a(\phi_i)$.
Here, we will content ourselves with describing
some features which will be important for model building, as well as some more general theoretical issues.

Perhaps most important for us will be the form of the scalar potential.
In units with $M_p = 1$, where $M_p$ is the reduced Planck mass,
\beq
M_p = {\sqrt{\hbar c \over 8 \pi G_N}}=  2.43 \times 10^{18} ~{\rm GeV}
\eeq
the scalar potential takes the form:
\beq
V = e^{K} \left [ D_i W
g^{i \bar i} D_{\bar i} W^*- 3 \vert W \vert^2 \right ].
\label{sugrapotential}
\eeq
$D_i \phi \equiv F_i$ is an
order parameter for SUSY breaking:
\beq
D_i W = {\partial W \over \partial \phi_i} + {\partial K \over \partial \phi_i} W.
\eeq
From the form of the potential, \eq{sugrapotential}, we see that
if supersymmetry is unbroken, space time is Minkowski if $W=0$, and AdS if $W \ne 0$.
If supersymmetry is broken, and the space-time is approximately flat space ($\langle V \rangle = 0$), then
\beq
m_{3/2} =\langle e^{K/2} W \rangle.
\eeq

\subsubsection{Supersymmetry as a Gauge Symmetry}

In a general sense, one cannot meaningfully speak of spontaneously breaking a gauge symmetry\cite{Fradkin:1978dv,'tHooft:1979bh}.   Gauge symmetries are
redundancies in the description of a system, and one can only speak sensibly about expectation values of
gauge invariant operators.   Supersymmetry, if a symmetry
of nature, is necessarily a local symmetry (at a a primitive level, one can simply
argue that if gravitational interactions are not supersymmetric, one expects huge radiative
corrections to superpartner masses).  This same principle would appear to apply to supersymmetry as well.  In particular, the statement that
supersymmetry is necessarily an element of string theory, and thus, at some scale, one expects evidence of supersymmetry, is not correct.  This point is well-illustrated
by various non-supersymmetric string constructions.

On the other hand, in an ordinary gauge theory, if the coupling is weak, the notion of broken gauge symmetry {\it is a useful one}.  In the limit of vanishing
 $g$, the symmetry is a global symmetry, and it is meaningful to speak of Goldstone bosons; for very
  small $g$, there is a range of scales at which the spectrum can sensibly be described in terms of Goldstone bosons and gauge bosons and, for finite momentum, the Goldstone bosons
are more strongly interacting than the gauge bosons (this is the essence of the Goldstone boson equivalence theorem).  For example, in a technicolor-like theory, in which
the $SU(2)$ gauge coupling is small, the theory exhibits approximate {\it global} symmetries, which are important
guides to model building and phenomenology.   A similar statement applies to supersymmetry, where now the issue
is the size of the quantity $G F$, where $G$ is Newton's constant and $F$ is the order parameter of supersymmetry breaking; for small values of this quantity, there is a light
field, the Goldstino, whose couplings are governed by low energy theorems.  When this parameter is large, there is no such statement, and no distinctive consequence
of the low underlying gauge symmetry\cite{Dine:2006ii}.  It should be noted that there are sharp arguments against arbitrarily small gauge couplings in theories
of gravity\cite{ArkaniHamed:2006dz}.

\subsubsection{Intermediate Scale (Gravity) Mediation}

Most work on the phenomenology of supersymmetry has been performed in the framework of ``gravity mediation", or perhaps more properly ``Intermediate Scale
Mediation".   The basic premise of these theories is that supersymmetry is broken in a hidden sector, typically by an O'Raifeartaigh-like model, and the breaking is transmitted
to the fields of the MSSM through Planck suppressed operators.  The ideas are illustrated by supposing that we have a field, $X$, with
\beq
F_X \equiv D_X W \ne 0.
\eeq
and with a Kahler potential for $X$ and matter fields, $\phi_i$, which is simply
\beq
K = X^\dagger X + \sum \phi_i^\dagger \phi_i.
\eeq
One also supposes that there is a constant in $W$, $W_0$, chosen so that the energy of the ground state is (nearly) zero.  This means that
\beq
\vert \langle W \rangle \vert = \sqrt{3} F_X,
\eeq
and
\beq
m_{3/2} = \langle e^{K/2} W \rangle.
\eeq
Then, for example, there is a term in the potential for the $\phi_i$'s:
\beq
V(\phi_i) = m_{3/2}^2 \vert \phi_i \vert^2.
\eeq
If we restore the factors of $M_p$,
\beq
m_{3/2} \approx {F_x \over M_p},
\eeq
so if we wish the scale of supersymmetry breaking to be of order $1$ TeV, we require $F_X \approx \sqrt{m_{3/2} M_p} \approx (10^{11} {\rm GeV})^2.$
This is the origin of the term ``intermediate scale."  We also see that, already in this simple model, all scalars gain mass of order $m_{3/2}$.  Additional
soft breakings readily arise as well:   cubic terms in the scalar potential, proportional to $m_{3/2}$ (so-called $A$ terms), and
gaugino masses.

We will not discuss models for the hidden sector field(s) $X$ in any detail, but it is interesting to note that different issues arise
than in the case of globally supersymmetric, O'Raifeartaigh models.  The equations $F_i = 0$ are no longer holomorphic.
In many existing models, fields have vev's of order $M_p$, and one cannot make general statements about
the requirements for susy-breaking minima.  If fields are small, perhaps due to discrete symmetries, than
the arguments of Nelson and Seiberg apply, at least approximately.

There is at least one troubling feature of intermediate scale models.  In our simple example, the squarks and sleptons are all degenerate.
This is a desirable feature; it leads to suppression of flavor changing processes such as $K^0 \leftrightarrow \bar K^0$ mixing.  But there is no
symmetry which enforces this.  Any such would-be flavor symmetry is broken by the quark and lepton Yukawa couplings.
A more complicated Kahler potential, including terms such as
\beq
\delta K = {1 \over M^2} \gamma^{ij} X^\dagger X Q_i^\dagger Q^j
\eeq
can lead to an almost random mass matrix for the squarks and sleptons.  Only if the coefficient of this term is very small or its structure
highly restricted can it be compatible with data from flavor physics.  This problem accounts for the appeal of gauge mediation, to be discussed
in the next section.

\subsection{Gauge Mediation}

The main premiss underlying gauge mediation can be simply described:  in the limit that the gauge couplings vanish, the hidden and visible sectors decouple.\footnote{This
definition was most clearly stated in \cite{Meade:2008wd}, but some care is required, since, as we will see, additional features are required for a realistic model.}

The simplest model of gauge mediation, known as Minimal Gauge Mediation (MGM) (for a review see \cite{Giudice:1998bp}), involves a chiral field, $X$, whose vacuum expectation value
is assumed to take the form:
\beq
\langle X \rangle = x + \theta^2 F.
 \eeq
 $X$ is coupled to a vector-like set of fields, transforming as $5$ and $\bar 5$ of $SU(5)$:
 \beq
 W = X(\lambda_\ell \bar \ell \ell + \lambda_q \bar q q).
\label{simplegm}
\eeq
For $F<X$, $\ell, \bar \ell, q, \bar q$ are massive, with supersymmetry breaking splittings of order $F$.
The fermion masses are given by:
\beq
m_q = \lambda_q x;~~~ m_\ell = \ \lambda_\ell  x,
\eeq
while the scalar splittings are
\beq
\Delta m_q^2 = \lambda_q F; ~~~~~ \Delta m_\ell^2 = \lambda_\ell F.
\eeq

 In such a model, masses for gauginos are generated at one loop; for scalars at two loops.  The gaugino mass computation
 is quite simple.  Even the two loop scalar
 masses turn out to be rather easy, as one is working at zero momentum.  The latter calculation
can be done quite efficiently using supergraph
techniques; an elegant alternative uses background field arguments\cite{Giudice:1997ni,ArkaniHamed:1998kj}.
The result for the gaugino masses is:
\beq
m_{\lambda_i} = {\alpha_i \over \pi} \Lambda,
\eeq
For the squark and slepton masses, one finds
\beq
\widetilde m^2 ={2 \Lambda^2}
[
C_3\left({\alpha_3 \over 4 \pi}\right)^2
+C_2\left({\alpha_2\over 4 \pi}\right)^2
+{5 \over 3}{\left(Y\over2\right)^2}
\left({\alpha_1\over 4 \pi}\right)^2 ],
\label{scalarsmgm}
\eeq
where $\Lambda = F_x/x$.
$C_3 = 4/3$ for color triplets and zero for singlets,
$C_2= 3/4$ for
weak doublets and zero for singlets.

\subsubsection{Features of MGM}

MGM has a number of features which make the model (really a class of models) interesting for phenomenology and model building.
\begin{enumerate}
\item  One parameter describes the masses of the three gauginos and the squarks and sleptons
\item  Flavor-changing neutral currents are automatically suppressed; each of the matrices $m_Q^2$, etc., is automatically proportional to the
unit matrix; the $A$ terms are highly suppressed (they receive no contributions before three loop order).
\item  CP conservation is automatic
\item  The strict definition of gauge mediation
given above fails at some level.  In particular, such models cannot generate a $\mu$ term; the term is protected by symmetries.  Some further structure is necessary.
\end{enumerate}

Much like the simplest supergravity modes, MGM has proven a useful model for understanding experimental limits, and it could be the ultimate,
underlying theory.  But there are a number of reasons to consider possible generalizations, as we will in the next section.

\subsection{General Gauge Mediation}

Much work has been devoted to understanding the properties of this simple model, but it is natural to ask:
just how general are these features?  It turns out that some are peculiar to our assumption of a single set of messengers
and just one singlet responsible for supersymmetry breaking and R symmetry breaking.
Meade, Seiberg and Shih have formulated the problem of gauge mediation in a general way,
and dubbed this formulation {\it General Gauge Mediation}  (GGM)\cite{Meade:2008wd}.    They study the problem
in terms of correlation functions of (gauge) supercurrents.  Analyzing the restrictions imposed by Lorentz invariance and supersymmetry
on these correlation functions, they find that the general gauge-mediated spectrum is described by three complex parameters and three real
parameters.  The models share the feature of the minimal models in that the masses of squarks, sleptons, and gauginos
are functions only of their gauge quantum numbers, so the flavor problems are still mitigated.  The basic features of the spectrum
are readily understood.  There are three independent Majorana masses for the gauginos (these are the three complex parameters
mentioned above).  The scalar mass spectrum is described by three real parameters, $\Lambda_i$:
\beq
\widetilde m^2 = 2
\left[
C_3\left({\alpha_3 \over 4 \pi}\right)^2 \Lambda_c^2
+C_2\left({\alpha_2\over 4 \pi}\right)^2 \Lambda_w^2
+{5 \over 3}{\left(Y\over2\right)^2}
\left({\alpha_1\over 4 \pi}\right)^2 \Lambda_Y^2\right].
\eeq
Ref. \cite{Meade:2008wd} relate each of these parameters to correlation functions of (generalized) currents.

One can construct models which exhibit the full set of possible parameters by including multiple chiral singlets
coupled to multiple messengers\cite{Carpenter:2008wi}.  Actually filling out the full parameter space is more challenging\cite{Buican:2008ws,Dumitrescu:2010ha}. Gauge messengers can also contribute to a general spectrum \cite{Intriligator:2010be}.
In any case, these generalized models show promise in solving some of the more troubling issues of gauge mediated
model building.  Most importantly, they can lead to a spectrum where squark and slepton
masses are comparable. In minimal gauge mediation, the known limits on the masses of the lightest sleptons imply that the
squarks are very heavy.  These, in turn, give large loop contributions to the Higgs masses, and require significant
fine tuning.  The ``compression" of the spectrum possible in GGM can ameliorate this problem.  However,
these models also pose new challenges; potential problems
arise due to the proliferation of parameters.  Probably most significant of that relative phases in the gaugino masses can
lead to unacceptably large electric dipole moments for the neutron and electron\footnote{We thank
Nima Arkani-Hamed for stressing this point.}.

\section{Dynamical Supersymmetry Breaking}

\subsection{Models with Stable DSB}

We are now prepared to deal with dynamical supersymmetry breaking.
Models with stable dynamical supersymmetry breaking are rare.  For reviews, see, for example, \cite{Shadmi:1999jy}.  They are generally characterized by two features\cite{Affleck:1984uz,Affleck:1983vc}:
\begin{enumerate}
\item  Classically, their potentials have no flat directions (there is not a moduli space of vacua).
\item  They exhibit global symmetries, which are spontaneously broken in the ground state.
\end{enumerate}
If the first condition is {\it not} satisfied, then, as we have seen, there are typically regions in the (pseudo-) moduli space where the potential
tends to zero, corresponding to (at least) asymptotic restoration of supersymmetry.  If the second condition is satisfied, there
are some number of Goldstone particles.  In general, as in the example of massless QCD, these particles each lie in a different
chiral multiplet.  The other scalar in the multiplet, like the Goldstone, will have no potential; it is a flat direction, contradicting the
first assumption above.  There is a potential loophole in this argument:  it is logically possible that both fields in the multiplet
are Goldstone particles.  No examples are known in which this phenomenon occurs, and in fact general arguments demonstrate
that it is inconsistent\cite{Komargodski:2010rb}.

We will content ourselves with a small number of examples; many more can be found in the texts and reviews cited.
The simplest example of such a theory, in which it is possible to do systematic calculations, is known
as the $3-2$ model because the gauge group is $SU(3) \times SU(2)$.  Its particle content is like that of
a single generation of the Standard Model, minus the singlet:
\beq
Q:(3,2)~~\bar U:(\bar 3,1) ~~\bar D:(\bar 3,1)~~L = (1,2).
\eeq
There is a unique renormalizable superpotential allowed by the symmetries, up to field redefinitions:
\beq
W = \lambda Q L \bar U.
\label{32w}
\eeq

This model satisfies both of our criteria above.
To see that there are no flat directions, consider first the theory with $\lambda =0$.
Without loss of generality,
one can take, for the flat direction of the $D$-term:

\beq
Q = \left ( \matrix{a & 0 & 0 \cr 0 & b & 0} \right )~~~~ L = \left ( \matrix{e^{i \phi} \sqrt{\vert a \vert^2 - \vert b \vert^2}\cr 0} \right )
\eeq
$$\bar U = \left ( \matrix{a & 0 & 0} \right )~~\bar D = \left ( \matrix{0 & b & 0} \right ).$$
Turning on $\lambda$, one cannot satisfy all of the ${\partial W \over \partial \phi_i} =0$ equations, unless both $a$ and
$b$ vanish.  Finally, using the techniques familiar from our discussion of supersymmetric
 QCD, the model possesses a non-anomalous $R$ symmetry which is spontaneously broken by a non-vanishing
$a$ or $b$.

Without the superpotential of \eq{32w}, and assuming that the $SU(2)$ coupling is much smaller than the $SU(3)$ coupling,
this is supersymmetric QCD with $N=3,N_f =2$.
The theory has a set of flat directions, and generates a non-perturbative superpotential.
For small $\lambda$, the effective superpotential is the sum of the (marginal) classical superpotential and
the (relevant) non-perturbative one:
\beq
W_{eff} = {\Lambda^6 \over Q Q \bar U \bar D} + \lambda Q L \bar U.
\eeq
Careful study of the resulting potential exhibits a supersymmetry-breaking minimum.

One can ask:  what happens in this model if the $SU(2)$ coupling is much greater than the $SU(3)$ coupling, so that the $SU(2)$ gauge group becomes strong first.
In this limit, the theory looks like $QCD$ with $N=2, ~N_f=2$.  In this theory, there is no non-perturbative superpotential:  there
exists an exact moduli
space, even quantum mechanically.  However, as Seiberg showed\cite{Seiberg:1994bz,Seiberg:1994pq} and we reviewed
in section \ref{nfequalsnc}, in such theories, the moduli space is modified quantum mechanically.
In effect, $QL$ is non-zero everywhere on the moduli space, generating an $F$ term for $\bar U$ through \eq{32w}\cite{Intriligator:1996fk} (for a pedagogical
discussion of the $3-2$ model, including this issue, see \cite{Dine:2007zp}, sections 14.1, 16.3.1).
Other models exploiting the quantum moduli space to break supersymmetry are presented in \cite{Izawa:1996pk,Intriligator:1996pu}

Finally, what if we give up the requirement of renormalizability?  Adding higher dimension operators (e.g. $(Q \bar U L)^2$), with
coefficients scaled by a large mass, $M$,  we will lose the
continuous $R$ symmetry as an exact symmetry, and there will be supersymmetric minima.  These minima, however, will be far away, and separated
by a large barrier (with barrier height scaled by $M$) from the non-supersymmetric minimum near the origin.  The metastable state near the origin will be extremely stable.

It is not easy to find theories which satisfy the conditions for stable supersymmetry breaking, and those which exist pose challenges for model building.
To illustrate the issues, we can consider a class of models with gauge group $SU(N)$, and an antisymmetric tensor, $A_{ij}$, as well as $N-4$ antifundamentals, $\bar F$.
The simplest of these models has $N=5$, and a single $\bar 5$.  It is easy to check, using the matrix technique developed above, that there are no
flat directions of the $D$ terms.   There is a non-anomalous $R$ symmetry; one can give arguments that, in this strongly coupled theory, the symmetry
is spontaneously broken\cite{Affleck:1983vc}.  These features extend to the model with general $N$, when we
include the most general superpotential
\beq
W = \lambda_{ab} \bar F^a_i \bar F^b_j A^{ij}.
\eeq
So we might adopt the following strategy for model building.  Take $N_f$ (and hence $N$) and choose $\lambda$ appropriately so that the model
has a large flavor group.  For example, for $N=14$, the flavor group can include $SU(5)$.  One can gauge a subgroup of the flavor group, and identify this subgroup with the
Standard Model gauge group.  Then loop effects can give rise to
masses for the fields of the MSSM.  This breaking is said to be direct, since the messengers sector and the supersymmetry
  breaking sector are one in the same.  The spectrum is not calculable, but falls within the rubric of GGM; masses of the
  MSSM fields can be expressed in terms of gauge couplings and certain
  correlation functions of the strongly coupled theory.

  The most serious difficulty of this model, and those like it, is that the QCD and other gauge couplings are violently non-asymptotically free.  Unification is lost;
 indeed, some enlargement of the Standard Model group is required only a few decades above the SUSY-breaking scale.  Most model building with
 stable supersymmetry breaking invokes more complicated structures in order to obtain a vev for a field like $X$ of the O'Raifeartaigh models, which in turn couples to messengers.
 The constructions are rather baroque\cite{Dine:1995ag,Dine:1993yw,Dine:1994vc}.

\subsection{Non-Linear Lagrangians}

In QCD, the breaking of the approximate chiral symmetries gives rise to Goldstone bosons, the $\pi$ mesons (more generally, the pseudoscalar meson
octet).  At low energies, these are the only active degrees of freedom, and one should integrate out the other hadrons.
The symmetries tightly constrain the dynamics of these fields; these constraints are most simply
incorporated  through the use of a non-linear effective lagrangian.  In this lagrangian, the basic degrees of freedom
can be taken to be (special) unitary matrices, $U$, which transform simply under the symmetry group:
\beq
U(x)  \rightarrow L U(x) R^\dagger,
\eeq
where $L$ and $R$ are matrices in $SU(2)_L$ and $SU(2)_R$.
The unitarity constraint is satisfied by parameterizing the fields as:
\beq
U(x) = e^{i {\pi^a(x) \tau^a  \over 2f_a}}.
\eeq
The terms with
two derivatives of the fields are unique, up to normalization:
\beq
{\cal L} = f_\pi^2 {\rm Tr} (\partial_\mu U^\dagger \partial^\mu U).
\eeq
Quite generally, such non-linear lagrangians are very useful for understanding the dynamics of strongly coupled theories.

For supersymmetry, one can proceed analogously\cite{Akulov:1974xz,Wess:1992cp,Komargodski:2009rz}.
We will follow \cite{Komargodski:2009rz}.  The first step is to determine the analog of the
constrained field $U$ of the non-linear sigma model.  It is helpful to think of O'Raifeartaigh models in deciding
how to proceed.  In this case, one has some fields which are relatively heavy, with mass $m$.  One has the
scalar pseudomodulus, $x$, which is lighter, a pseudoscalar, $a$, which is either degenerate with $x$, or (in the case of spontaneous
breaking of the $R$ symmetry), massless, and the Goldstino, which is strictly massless.
Let's focus, first, on the case where the axion is also massive.  Our goal, then, is to describe the system
at very low energies, which consists purely of the Goldstino.   Let's start with the effective action for the full
chiral field $X$, at scales below $m$ but above the mass of $x$.  This has the form.
\beq
{\cal L} = \int d^4 \theta \left (X^\dagger X + {c \over m^2} X^\dagger X X^\dagger X \right ) + \int d^2 \theta f X.
\eeq
The lagrangian for the component fields, calling $G$ the Goldstino, reads:
\beq
{c \over m^2} \left (4 f^\dagger f x^\dagger x + 2 x f G^{*2}+ {\rm c.c.}+ \dots \right )
\eeq
so, solving the $x$ equation of motion, we have for the $X$ {\it superfield}:
\beq
X = {G^2 \over 2f} + \sqrt{2} \theta G + \theta^2 F.
\eeq
As a result, $X$ obeys the {\it supersymmetric} constraint:
\beq
X^2 =0.
\eeq
Substituting this into original lagrangian yields the low energy lagrangian for $G$, analogous to the low energy pion lagrangian:
\beq
{\cal L}_{AV} = -f^2 + i \partial_\mu G^* \sigma^\mu G + { 1\over 4 f^2} G^{*2} \partial^2 G^2 - {1 \over 16 f^6} G^2 \bar G^2 \partial^2 G^2 \partial^2 \bar G^2.
\eeq
This is the Akulov-Volkov lagrangian\cite{Akulov:1974xz}, one of the first implementations of supersymmetry.  While observed for this particular model, the form
of the constraint and the non-linear lagrangian, as in the pion case, are quite general.

This treatment can be modified to include other massless fields, particularly the axion which results if the $R$ symmetry is spontaneously
broken\cite{Komargodski:2009rz}.  Apart from applications to the interactions of Goldstinos, other constraints on supersymmetric theories can be obtained from this
type of analysis.  In \cite{Dine:2009sw}, the non-linear lagrangian plays an important role in proving a bound on the superpotential:
\beq
W < {1\over 2} f_a f,
\eeq
where $f_a$ is the $R$-axion decay constant, and $f$ is the Goldstino decay constant.  Such lagrangians
 have been used to constrain ${\mathcal N} = 2$ theories \cite{Antoniadis:2010nj}; potential effects on phenomenology were discussed in \cite{Antoniadis:2010hs}.

\section{Metastable Supersymmetry Breaking}

We have remarked that stable supersymmetry breaking is special, and model building incorporating it complicated.  Metastable supersymmetry breaking turns out to
be much more generic, and greatly expands the possibilities for model building.
But there is a more principled reason why metastability might be expected to be important in supersymmetry breaking, and this is connected with the Nelson-Seiberg
theorem\cite{Nelson:1993nf}.  Generic supersymmetry breaking requires the existence of a continuous $R$ symmetry.  In theories containing general relativity (i.e. the theory which
describes the world around us), it is generally believed that there can be no continuous global symmetries.  In the theories with stable supersymmetry breaking, we might
expect the appearance
of supersymmetric ground states as a result of small, $R$ symmetry breaking effects, perhaps through higher dimension operators.  So some degree of metastability
appears inevitable.

Intriligator, Seiberg, and Shih (ISS) \cite{Intriligator:2006dd} provided an  example of metastable supersymmetry breaking in a surprising setting:  vectorlike supersymmetric
QCD (perhaps the earliest suggestion that metastable supersymmetry breaking might be important
appears in \cite{Ellis:1982vi}).  At a broader level, brought their work leads to the realization that metastable supersymmetry breaking is a generic phenomenon\cite{Intriligator:2007cp}.  Consider
the Nelson-Seiberg theorem, which asserts that, to be generic, supersymmetry breaking requires a global, continuous $R$ symmetry.
We expect that such symmetries are, at best, accidental low energy consequences of other features of some more microscopic theory.
So they will be violated by higher dimension operators, and typical theories will exhibit SUSY-preserving ground states.
In this section, we first illustrate the issues within the context of O'Raifeartaigh models perturbed by higher dimension operators.
A crucial role for discrete $R$ symmetries will be apparent in this context.  We then turn to the ISS model, explaining how, in this strongly
  coupled theory, one can establish the existence of a metastable, non-supersymmetric state.  We elucidate a number of features of the model
  relevant to its phenomenology.  In section \ref{retrofitting}, we study ``Retrofitted" models, a general class of models exhibiting dynamical, metastable supersymmetry
breaking, which allow rich possibilities for model building.

\subsection{Metastability in Supersymmetric Quantum Mechanics and Classical Supersymmetric Field Theories}

Perhaps the simplest setting in which to consider the question of metastability is supersymmetric quantum mechanics.  We can easily choose functions $W$ in \eq{susyqm} such that the potential has a local minimum which is unstable to decay to a supersymmetric minimum.
By arranging the barrier to be sufficiently high, we can make the lifetime of this metastable arbitrarily long.  This might serve as a plausible model
for supersymmetry breaking in our universe.

On the other hand, in classical supersymmetric field theories, with Kahler potentials which are simply quadratic (canonical) in all fields,
one cannot obtain such a metastable state\cite{Ray:2006wk,Sun:2008va}.  In particular, this means that one cannot obtain metastability at tree level
in renormalizable theories.  If we allow more general Kahler potentials, it is clear that metastability is possible in principle.  As we will see in the following sections, non-renormalizable theories, after the inclusion of quantum effects, and a variety of strongly coupled quantum theories do allow the possibility of metastability.
We will argue that if our universe does exhibit low energy supersymmetry, we are likely living in such a metastable state.

\subsection{Continuous Symmetry from a Discrete Symmetry}

The continuous symmetry of the O'Raifeartaigh model might arise as an accidental consequence of a discrete, $Z_N$ {\cal R} symmetry.
This could simply be a subgroup of the original $R$ symmetry.  For example:
\beq
X \rightarrow e^{4 \pi i \over N} X;~Y \rightarrow e^{4 \pi i \over N} Y
\eeq
corresponding to $\alpha = {2 \pi \over N}$ in \eq{ror} above.

For general $N$, this symmetry is enough to ensure that, keeping only renormalizable terms,
the lagrangian is that of \eq{simplestor}.  But higher dimension terms can break
the continuous $R$ symmetry.
Suppose, for example, $N=5$.  The discrete symmetry now allows couplings such as
\beq
\delta {W} = {1 \over M^3} \left ( a X^6 + b Y^6 + c X^4 Y^2 + d X^2 Y^4 + \dots \right ).
\eeq
Note that $W$ transforms, as it must, under the discrete $R$
symmetry,  $W \rightarrow e^{4 \pi i \over N} W$.

The theory now has $N$ supersymmetric minima, with
\beq X \sim \left ( \mu^2 M^3 \right )^{1/5} \alpha^k,
\eeq
where $\alpha = e^{2 \pi i \over 5}$, $k = 1,\dots,5$, corresponding to spontaneous
breaking of the discrete symmetry.  Note that for large $M$, this vacuum is ``far away"
from the origin;
the theory still has a supersymmetry breaking minimum near the origin.  Indeed, near the origin, the
higher dimension (irrelevant) operator has no appreciable effect, and the Coleman-Weinberg calculation
goes through as before.

\subsection{Metastability}

The broken supersymmetry state near the origin, at least in the limit of global supersymmetry, will eventually
decay to one of the supersymmetric minima far away.
We can ask how quickly this decay occurs.  We will content ourselves with simple scaling arguments
for the example above, establishing a lower bound on the lifetime, which will turn out to be {\it extremely}
large.
In the semiclassical approximation\cite{Coleman:1977py}, one is instructed to construct a solution of the Euclidean
field equations, the ``bounce".  It is enough to look for $O(4)$ invariant solutions.
For a single field, $\phi$, one studies the equation:
\beq
\ddot \phi + {3 \over r} \dot \phi = V^\prime(\phi),
\label{bounceequation}
\eeq
subject to certain boundary conditions.
This is identical to the equation of motion for a particle in a potential $-V$, where $r$ plays the role of time, and the second
term is acts like a damping term.  One solves this equation, subject to the boundary conditions that the ``particle"
starts, with nearly zero velocity, in the ``true" vacuum, at time zero, and end up in the ``false" vacuum
at $r = \infty$.  From this solution, one constructs the (field theory) action:
\beq
S = \int dr r^3 \left ( {1 \over 2} \dot \phi^2 - V(r) \right ).
\eeq
The inverse lifetime (per unit volume) is
\beq
\Gamma = m^4 e^{-S},
\eeq
where $m$ is some characteristic energy scale (more precisely, one must evaluate a functional determinant, but this will not be important
for our considerations).

In our example above, the role of $\phi$ is played by $X$.  $\phi$ ($X$) changes by an amount of order
\beq
\Delta \phi = \mu \left ( {M \over \mu} \right )^{3/5}.
\eeq
The height of the potential barrier is of
order $\mu^4 \log(M)$.  If one ignores friction, one can solve \eq{bounceequation} by quadrature.  The action is then seen to scale as
\beq
S \approx \left ( {M \over \mu} \right )^{18/5}.
\eeq
This is a lower bound, as the effects of friction will only increase the action.  We have in mind situations where $M$ is many orders of magnitude larger than
$\mu$, so the lifetime is typically vastly larger than the age of the universe.




\subsection{ISS}

Intriligator, Seiberg and Shih, \cite{Intriligator:2006dd}, studied supersymmetric QCD with massive quarks and made a remarkable discovery.  On the one hand, from the Witten
 index, we know that all such theories have $N_c$ supersymmetric vacua.  However, for a particular range of $N_f$, one can show that the theories
 possess metastable, non-supersymmetric vacua in the regime of strong coupling.
  Specifically, they analyzed $SU(N_c)$ SQCD (having a dimensional transmutation scale ``$\Lambda$") with $N_c$ colors and with $N_f$ pairs of vector-like quarks ($Q$ and $\tilde{Q}$) in the IR-free window: $N_c <  N_f  < \frac{3}{2} N_c$, when the theory contains the non-zero superpotential
\beq
W^{(el)} = m \delta_{ij} \tilde{Q}^iQ^j.
\eeq

The duality of \cite{Seiberg:1994pq} suggests that the IR of the zero mass theory at or near the origin of its moduli space can be described by an $SU(N)$ ($N = N_f -N_c$) gauge theory (having a Landau pole at ``$\tilde{\Lambda}$") with $N_f$ pairs of vector-like quarks ($q$ and $\tilde{q}$) and a singlet ``meson" that transforms as a bi-fundamental under the global flavor symmetry of the model. Turning on the mass term as a small perturbation should not dramatically alter this picture near the origin. In the theory with a small mass the superpotential of the ``magnetic" dual becomes
\beq
\label{magsup}
W^{(mag)} = \frac{1}{\hat{\Lambda}}\tilde{q}_iM^{ij}q_j + m \delta_{ij}M^{ij}.
\eeq
with Kahler potential
\beq
K = \frac{1}{\beta}\left( \tilde{q}^{\dagger}\tilde{q} + q^{\dagger}q \right) + \frac{1}{\alpha|\Lambda|^2}{\rm Tr}\left( M^{\dagger}M \right),
\eeq
where
\beq
\Lambda^{3N_c -N_f} \tilde{\Lambda}^{3(N_f -N_c) -N_f} = (-1)^{N_f -N_c}\hat{\Lambda}^{N_f}.
\eeq
One can make the quark fields canonical, absorbing the coefficient $\beta$ into the definition of $\hat{\Lambda}$:$~~\hat{\Lambda}' = \frac{\hat{\Lambda}}{\beta}$. Similarly one can redefine the meson as $M \rightarrow \sqrt{\alpha}\Lambda \Phi$. That Kahler potential is then canonical,
\beq
K = \left( \tilde{\phi}^{\dagger}\tilde{\phi} + \phi^{\dagger}\phi \right) + {\rm Tr}\left( \Phi^{\dagger}\Phi \right).
\eeq
In this case the superpotential becomes
\beq
W = \frac{\sqrt{\alpha}\Lambda}{\hat{\Lambda}}\tilde{\phi}\Phi\phi + \sqrt{\alpha}m\Lambda {\rm Tr}\left( \Phi \right).
\eeq
Finally, defining $h = \frac{\sqrt{\alpha}\Lambda}{\hat{\Lambda}}$, and $\mu^2 = -m\hat{\Lambda}$, one converts the superpotential to:
\beq \label{magtheory}
W = h\tilde{\phi}\Phi\phi - h\mu^2 {\rm Tr}\left( \Phi \right)
\eeq
so that the dual theory is known up to two parameters, $\mu$ and $h$ (equivilently $\alpha$ and $\hat{\Lambda}'$).

For the specific case of $N_f = N_c + 1$, the theory s-confines so that there is no dual gauge group and an additional term in the superpotential is present, indicating the presence of a supersymmetric vacuum. We already know the superpotential in terms of the moduli of the electric theory,
\beq \label{sconf}
W^{(mag)}_{N_f = N_c +1} &=&\frac{1}{\Lambda^{2N_c -1}}( \tilde{B}_iM^{ij}B_j - det(M))+ m \delta_{ij}M^{ij}.
\eeq
The Kahler potential in this case is
\beq
K = \frac{1}{\beta}\left( \tilde{B}^{\dagger}\tilde{B} + B^{\dagger}B \right) + \frac{1}{\alpha|\Lambda|^2}{\rm Tr}\left( M^{\dagger}M \right).
\eeq
We may relabel: $B = \sqrt{\beta}\phi$, $\tilde{B} = \sqrt{\beta}\tilde{\phi}$, and $M = \sqrt{\alpha}\Phi$ and define $h = \frac{\beta\sqrt{\alpha}}{\Lambda^{2N_c-2}}$ and $\mu^2 = -\frac{m \Lambda^{2N_c-1}}{\beta}$. This converts the $N_f = N_c +1$ theory into the the theory of \eq{magtheory} with canonical superpotential. Again the theory is known up to two unknown parameters $h$ and $\mu^2$ (equivilently $\alpha$ and $\beta$).

 The superpotential of \eq{magtheory}, possesses the same global symmetries as the electric theory, $SU(N_f)_D \times U(1)_B \times U(1)_R$. In Tables \ref{sqcdEnm}, \ref{sqcdE}, \ref{sqcdMnm}, and  \ref{sqcdM} we list the symmetries of both massive and massless electric and magnetic theories and how the symmetries manifest themselves. Note that the $SU(N_f)_D$ is the diagonal subgroup of $SU(N_f)_L\times SU(N_f)_R$ that remains unbroken by the mass term.

\begin{table}[t!]
\centering
\begin{tabular}{|c||c|c|} \hline
 & $\quad Q \quad$ & $\quad \tilde{Q} \quad$  \\ \hline
 $SU(N_f)_L$ & $N_f$ & ${\bf 1}$ \\ \hline
 $SU(N_f)_R$ &${\bf 1}$ &  $\bar{N_f}$ \\ \hline
 $U(1)_B$ &$1$ &  $-1$ \\ \hline
 $U(1)_R$ & $1-\frac{N_c}{N_f}$ & $1-\frac{N_c}{N_f}$ \\ \hline
\end{tabular}
\caption{\label{sqcdEnm} Electric Theory no mass}
\end{table}

\begin{table}[t!]
\centering
\begin{tabular}{|c||c|c|c|} \hline
 & $\quad \phi \quad$ & $\quad \tilde{\phi} \quad$ & M \\ \hline
 $SU(N_f)_L$ & $\bar{N_f}$ & ${\bf 1}$ & $N_f$\\ \hline
 $SU(N_f)_R$ &${\bf 1}$ &  $N_f$ & $\bar{N_f}$ \\ \hline
 $U(1)_B$ & $\frac{N_c}{N_f-N_c}$ & $-\frac{N_c}{N_f-N_c}$ & 0 \\ \hline
 $U(1)_R$ & $\frac{N_c}{N_f}$ & $\frac{N_c}{N_f}$ & $2(1-\frac{N_c}{N_f})$ \\ \hline
\end{tabular}
\caption{\label{sqcdMnm} Magnetic Theory, no mass}
\end{table}

\begin{table}[t!]
\centering
\begin{tabular}{|c||c|c|} \hline
 & $\quad Q \quad$ & $\quad \tilde{Q} \quad$  \\ \hline
 $SU(N_f)_D$ & $N_f$ & $\bar{N_f}$ \\ \hline
 $U(1)_B$ &$1$ &  $-1$ \\ \hline
 $U(1)_R$ & $1$ & $1$ \\ \hline
  $Z_{2N}$ & $\alpha$  &  $\alpha$  \\ \hline
    $Z'_{2N}$ &$\alpha^2$  & $\alpha^0$  \\ \hline
\end{tabular}
\caption{\label{sqcdE} Electric Theory with mass: Here $\alpha = e^{-i\pi n/N_c}$ }.
\end{table}

\begin{table}[t!]
\centering
\begin{tabular}{|c||c|c|c|} \hline
 & $\quad \phi \quad$ & $\quad \tilde{\phi} \quad$ & M \\ \hline
 $SU(N_f)_D$ & $\bar{N_f}$ & $N_f$ & $N_f$ + $\bar{N_f}$\\ \hline
 $U(1)_B$ & $\frac{N_c}{N_f-N_c}$ & $-\frac{N_c}{N_f-N_c}$ & 0\\ \hline
 $U(1)_R$ & $0$ & $0$ & $2$ \\ \hline
  $Z_{2N}$ &$\alpha^{\frac{N_c}{N_f-N_c}}$ &  $\alpha^{-\frac{N_c}{N_f-N_c}}$& $\alpha^2$ \\ \hline
   $Z'_{2N}$ &$\alpha^0$ & $\alpha^0$ & $\alpha^2$  \\ \hline
  \end{tabular}
\caption{\label{sqcdM} Magnetic Theory with mass: Here $\alpha = e^{-i\pi n/N_c}$}.
\end{table}



In both massive theories the $U(1)_R$ is anomalous. However, the $Z_{2N}$ and $Z'_{2N}$ denote non-anomalous discrete symmetries left unbroken by the mass term in the electric theory. These discrete symmetries are anomaly free because they can be written as a linear combination of subgroups of  continuous anomaly free symmetries of the massless theory: The center of $SU(N_f)_L~~ ({\bf L})$, $U(1)_R ~~({\bf R})$, and $U(1)_B ~~({\bf B})$. In particular, $Z_{2N}$ \cite{Affleck:1983mk} is given by
\beq
Z_{2N} = {\bf L.R.B}; ~~~~~~~~~~~~~~~~~~~~~~\\{\bf L} = e^{-\frac{i 2\pi n}{N_f}}, ~~~~ {\bf R} =  e^{-\frac{i \pi n}{N_c}q_R}, ~~~~ {\bf B} = e^{\frac{i\pi n}{N_f}q_B},
\eeq
and $Z'_{2N}$ is given by
\beq
Z'_{2N} = {\bf L.R.B};~~~~~~~~~~~~~~~~~~~~~~\\{\bf L} = e^{-\frac{i 2\pi n}{N_f}}, ~~~~ {\bf R} =  e^{-\frac{i \pi n}{N_c}q_R}, ~~~~ {\bf B} = e^{-i\pi n\left(\frac{N_f-N_c}{N_cN_f}\right)q_B}.
\eeq
The $Z'_{2N}$ makes manifest that a magnetic quark vev will leave an unbroken $Z'_{2N}$ in the ISS vacuum.

In addition to the exact (non-anomalous) symmetries listed above, both theories have an approximate continuous $R$-symmetry, under which $R(\tilde{Q}Q) = R(M^{ij}) = 2$, and all other fields have $R$-charge zero.

With the dual theory in hand, we can analyze the vacua near the origin, \eq{magsup}. The $F$-term equations read:
\beq \label{rankeq}
\frac{\partial W}{\partial \Phi^{ij}} &=& \tilde{\phi}_i\phi_j - \mu^2 \delta_{ij} = 0,
\eeq
\beq
\frac{\partial W}{\partial \tilde{\phi}_i} &=& \Phi^{ij}\phi_j = 0,
\eeq
\beq
\frac{\partial W}{\partial \phi_j} &=& \tilde{\phi}_i\Phi^{ij} = 0.
\eeq
For $N<N_f$ the first equation, \eq{rankeq}, has no solution and SUSY is broken. The matrix $\tilde{q}_i^{\alpha} q_{j \alpha}$, with $i,j$ flavor indices and $\alpha$ color indices, has maximal rank $N$ due to the fact that it is essentially a sum of $N$ outer products of pairs of $N_f$ dimensional vectors. Since the rank of $\delta_{ij}$ is $N_f$, SUSY is broken by the ``rank condition."

The minimum energy configuration is acheived when the field vev's arrange themselves such that as many components of \eq{rankeq} are satisfied as possible. In other words:

\beq \label{ISSvac}
\Phi = \left( \begin{array}{cc} 0 & 0 \\  0 & \Phi_0
	           \end{array} \right),
	 ~~~ \phi = \left( \begin{array}{c} \phi_0 \\  0
	            \end{array} \right),
	~~~ \tilde{\phi}^T= \left( \begin{array}{c} \tilde{\phi}_0 \\ 0
		           \end{array}	   \right),         	
\eeq
with $\phi_0$ and $\tilde{\phi}_0$ each $N \times N$ matrices and $\Phi_0$ an $(N_f-N) \times (N_f -N)$ matrix.
The vacuum with the maximal amount of unbroken symmetry is the one with $\Phi_0 = 0 $ and $\tilde{\phi}_0 = \phi_0 = \mu {\bf 1}_N$. In this vacuum,
 many of the global symmetries are spontaneously broken:
\beq \label{Intriligator:2006ddglobalsymm}
& & G \rightarrow H,  \\
& & SU(N) \times SU(N_f) \times U(1)_B \times U(1)_R   \rightarrow  \\ & & SU(N)_D \times SU(N_f-N) \times U(1)_B' \times U(1)_R.  \nonumber
\eeq

Here $SU(N)_D$ is the diagonal subgroup of $SU(N) \times SU(N)_F$, where $SU(N)_F$ is a subgroup of the original $SU(N_f)$ group.  Counting the broken generators in G/H, yields $2 N N_f -N^2$ goldstone bosons. In addition to the massless goldstone degrees of freedom one also has pseudo-moduli directions. There are $2(N_f -N)^2 + N^2$ of these. The fact that the $SU(N)$ is gauged means that $N^2 -1$ of the goldstone bosons and $N^2 -1$ of the pseudo moduli are eaten. This leaves $2N(N_f-N) + 1$ goldstone directions and $2(N_f -N)^2 + 1$ pseudo moduli directions. $2(N_f-N)^2$ pseudo-moduli directions are perturbations about $\Phi_0$, while the one additional direction is a perturbation of $\phi_0+\tilde{\phi}_0^{\dagger}$. At one-loop these fields acquire positive squared mass:
\beq
 m^2_{\delta \Phi_0} &=& |h^4\mu^2| \frac{\log{4} -1}{8\pi^2}N, \\
 m^2_{(\delta (\phi_0-\tilde{\phi}_0))} & =&  |h^4\mu^2| \frac{\log{4} -1}{8\pi^2}(N_f-N).
\eeq
In general there are higher order terms in the Kahler potential which are not calculable and these contribute to the pseudo-moduli masses at order: $\mathcal{O}$$(\frac{|\mu^2|^2}{|\Lambda|^2})$. However, these are suppressed with respect to the one-loop masses if $\frac{h^4}{16\pi^2} > \frac{\mu^2}{|\Lambda|^2}$.  So, these vacua are calculable and locally stable near the origin.

Globally, however, one expects that SUSY preserving vacua exist due to the non-zero Witten index\cite{Witten:1982df} of pure SUSY Yang-Mills Theory. These SUSY preserving vacua appear at large values of the meson vev. One can see them by first giving the meson a large vev and then integrating out the dual quark matter. The theory without the quarks is simply pure SUSY Yang-Mills, but with a gaugino condensate, whose strength is dependent on the meson vev. This manifests itself as an additional meson dependent term in the superpotential of the effective theory:

\beq
W = N(h^{N_f} \Lambda^{3N-N_f}det(\Phi))^{\frac{1}{N}} + h\mu^2 Tr\Phi.
\eeq

As expected from the index,
the field equations of this effective theory have $N_c$ solutions:
\beq
\mu \ll \langle h\Phi \rangle = \Lambda \epsilon^{2N/(N_f-N)}{\bf 1}_{N_f} \ll \Lambda,
\eeq
where $\epsilon = \frac{\mu}{\Lambda}$.  We see that the
non-supersymmetric states near the origin are metastable, but their lifetime will be exponentially small in inverse powers of $\epsilon$. This can be showns from the exact tunneling solutions computed in \cite{Duncan:1992ai}.

\subsection{Gauge Mediation with ISS}

In the previous section we have reviewed two scenarios that realize SUSY breaking in a metastable state. Here we discuss developments in model building within this framework. We will see that each method has its challenges.

It is tempting to use the ISS model as a model of gauge mediation. Gauge mediation can be realized in a modular form where the ``messengers" are an extra ingredient, or it can be realized in a direct form where the role of the messenger fields are played by some part of the SUSY breaking sector. When attempting to use the ISS model for direct gauge mediated SUSY breaking, obstacles arise which prevent one from obtaining an acceptable superparticle spectrum. These require one to modify the basic ISS model.  In this section we review these issues and discuss some of the model building strategies which have been used to overcome these challenges. Many of these problems are encountered in the
older versions of gauge mediation.

The first difficulty is the vanishing of gaugino masses.  This arises as a result of an exact discrete $R$-symmetry which is preserved in the model (the $Z'_{2N}$ of Table \ref{sqcdM}). One requires that the $R$-symmetry be broken, either explicitly
or spontaneously. While explicit breaking can introduce nearby SUSY minima, spontaneous breaking can be difficult to arrange. An issue which arises specifically when using these models for direct mediation is that gaugino masses, while non-zero, are often suppressed at leading order in the $\frac{F}{M^2}$ expansion. The major distinction between most models is the mechanism by which they break $
 R
 $-symmetry and generate gaugino mass.

While the problem of $R$ symmetry breaking is the most serious for model building, there are others.  We will see that these do not preclude building phenomenologically
viable theories.
Using ISS for direct mediation of SUSY breaking typically ruins perturbative unification.
Due to the  presence of extra matter charged under the Standard Model gauge group, there are usually
Landau poles in the gauge coupling constants below the GUT scale, unless one chooses a high scale for
the messengers.  This typically requires that the models have a separation of scales between the messenger mass scale and SUSY breaking scale.  If the $F$-term sets the scale of SUSY breaking, then a new scale must be introduced into the theory to set the messenger mass.

The small ISS mass parameter itself may not be troubling. The stability and calculability of the ISS vacuum requires that $\frac{m}{\Lambda} \ll 1$. If one requires that all mass scales arise dynamically, then the smallness of the mass term requires explanation. One may use a retro-fitting procedure to dynamically explain the smallness of the mass term, as in section \ref{retrofitting}.  But one is still left with one or more remarkable coincidences of scale.

\subsubsection{Models with Spontaneous $R$-symmetry Breaking}

One simple strategy for building a model of gauge mediation within the framework provided by the ISS model of SUSY breaking is to realize direct gauge mediation by gauging a subgroup of the global flavor symmetry group. The authors of \cite{Csaki:2006wi} considered the case of $N_f = N_c +1$, ``s-confining" SQCD.  For $N_f =6$. this theory breaks SUSY and leaves over an unbroken $SU(5)$ global flavor group. Gauging this $SU(5)$ leads to soft masses for the MSSM scalars, but leaves the gauginos massless. In order to remedy this one must introduce explicit or spontaneous $R$-symmetry breaking. Explicit breaking tends to introduce operators which lead to
 supersymmetric vacua, and these must be forbidden.  Adding a new $U(1)$ (as in \cite{Dine:2006xt}) under which  new fields $Z,S (\bar{Z}, \bar{S})$ have charges $+1 (-1)$ with superpotential,
\beq
W = Tr(M)S\bar{S} + m(S\bar{Z} + Z \bar{S}),
\eeq
where $Tr(M)$ is the trace of the gauged components of the meson field. This leads to spontaneous $R$-symmetry breaking and a vev for the meson. In general the leading order contribution (in $\frac{F}{\langle M \rangle}$) to the gaugino mass is given by the formula \cite{Cheung:2007es}:
\beq
m_{\lambda} = \frac{g^2}{16 \pi^2} F_{X} \frac{\partial~ {\rm Log} \left( {\rm det}[{\mathcal M}(X)]\right) }{\partial X}
\eeq
where ${\mathcal M}(X)$ is the mass matrix of the messengers as a function of the field $X$ whose $F$-term breaks supersymmetry. This formula arises due to the dependence of the gauge coupling on the scale at which  heavy charged fields with SUSY breaking mass splittings are integrated out. In Minimal Gauge Mediation ${\rm Tr}(M) = X$,
and $M$ is the supersymmetry conserving part of the mass of the messenger fields. For this model, however,
\beq
{\rm det}({\mathcal M}(X)) = h^2\mu^2.
\eeq
As a result, gaugino masses vanish to leading order in $F/\langle X \rangle$, but are generated at ${\mathcal O}(\frac{F^3}{\langle X \rangle^5})$. The gaugino masses vanish
 more generally at leading order because the SUSY breaking vacuum in the  effective theory at the renormalizable level is the absolute ground state of the system \cite{Komargodski:2009jf}. This higher order generation of gaugino mass is an issue in any model that breaks SUSY in a global minimum.

 A different strategy to achieve spontaneous $R$-symmetry breaking was utilized in \cite{Abel:2007jx,Abel:2007nr}. Here one includes an additional (small) perturbation to the $N_f = N_c -2 = 7$ electric theory. The electric theory is:
\beq
W^{(el)} = m\delta_{ij}\tilde{Q}^i Q^j + \frac{1}{\Lambda^2_{UV}}(Q^5)^{i_1i_2}\epsilon_{i_1i_2}.
\eeq
where $i,j$ run over ${\emph only~two}$ flavor indices. This extra term leaves invariant an $SU(2) \times SU(5)$ of the diagonal $SU(7)$ flavor group. The resulting magnetic theory is:
\beq
W^{(mag)} =h\tilde{\phi}_i\Phi^{ij}\phi_j -\mu^2 \delta_{ij} \Phi^{ij} + m \epsilon_{\alpha \beta}\epsilon^{ij}\phi^{\alpha}_i\phi^{\beta}_j,
\eeq
where $m = \frac{\Lambda^3}{\Lambda_{UV}^2}$.
Gauging the global $SU(5)$ under the SM gauge group results in direct mediation of SUSY breaking. In this case spontaneous $R$-symmetry breaking is automatic. This is due the theorem of \cite{Shih:2007av} which we encountered earlier: A field with $R$-charge different than 0,2 is necessary for the origin to be destabilized by the one-loop Coleman-Weinberg potential. Here the first and second terms demand that $R(\phi_a) = -R(\tilde{\phi}_a)$ and the third term demands that $R(\phi_1) + R(\tilde{\phi}_2) = 2$. At least one field must have $R \not= 0,2$ in order for these conditions to be satisfied, thus the necessary condition for spontaneous $R$-symmetry breaking is satisfied. An explicit computation reveals that vacua with spontaneously broken $R$-symmetry are locally stable.

In \cite{Cho:2007yn} and \cite{Abel:2008tx} the theorem of \cite{Shih:2007av} is implemented by adding singlet fields which are forced to have $R$ charges $ \not= 0,2$. The deformation of \cite{Cho:2007yn} is:
\beq
\delta W = AB{\rm Tr}(\Phi) + mA^2.
\eeq
As expected, this model can have vacua with spontaneously broken $R$-symmetry.

Since $R$-symmetry is broken, the gaugino masses will generally be non-zero. An explicit computation reveals that ${\rm det}({\mathcal M}(X)) = {\rm const}$, where $X$ represents the field carrying non-zero $F$-term component. So, the leading order contribution to gaugino masses vanish, but are non-zero at order ${\mathcal O}(\frac{F^3}{M^5})$.

The basic challenge to arranging $R$ symmetry breaking is that the one-loop Coleman-Weinberg potential wants to stabilize flat directions at the origin rather than destabilizing them there. In \cite{Giveon:2008wp}, it was shown that some pseudo-moduli remain massless at one-loop if there are massless flavors in an ISS model. In this case the fate of these pseudo-moduli are determined by a two-loop computation. It is possible for this contribution to be negative, destabilizing the $R$-preserving point. This fact was used in \cite{Giveon:2008ne} to incorporate spontaneous $R$-symmetry breaking automatically in an ISS model, which was then used for gauge mediation. This strategy starts by considering the ISS model with $N_f = N_{f_0} + N_{f1}$ with $N_{f_0}$ massless and $N_{f1}$ massive flavors plus a higher order interaction for the massless flavors.
\beq \label{Giveon}
W = m_1 \delta_{\alpha \beta} \tilde{Q}^{\alpha}Q^{\beta} + \delta_{ab} \frac{\left( \tilde{Q}^aQ^c\tilde{Q}_cQ^b \right)}{M^*},
\eeq
with $\alpha, \beta = 1,..,N_{f1}$ and $a,b = 1,..,N_{f0}$.
The flavor symmetry of the model is:
\beq \label{globsymmks}
 SU(N) \times SU(N_{f0}) \times SU(N_{f1}) \times U(1)^3 \times U(1)_R.
 \eeq
As long as $N_{f1} > N_f -N_c$, SUSY will be spontaneously broken. The dual theory is just that of \eq{magtheory} with the additional term:
\beq
\delta W = \frac{1}{2}h\epsilon \mu {\rm Tr}(\Phi_{00}^2), ~~~~~~~~~~ \epsilon \sim \frac{\Lambda}{M_*}.
\eeq

  Expanding about the maximally symmetric vacuum ( \eq{ISSvac} with $\Phi_0 = 0$ and ($\phi = \tilde{\phi} = \mu {\bf 1}_{Nf1}$) one finds that some of the pseudomoduli fields do not receive masses at one-loop. Reorganizing the $(N_f -N) \times (N_f -N)$ matrix of meson pseudo-moduli ($\Phi_0$ in \eq{ISSvac}) into $\Phi_{11}$ (an $N_{f1} \times N_{f1}$ field), $\Phi_{10}$ and $\Phi_{01}^T$ ($N_{f1} \times N_{f0}$ fields ), and $\Phi_{00}$ (an $N_{f0} \times N_{f0}$ field):

 \beq
  \Phi_0 = \left( \begin{array}{cc} \Phi_{11} & \Phi_{01} \\  \Phi_{10} & \Phi_{00}
	           \end{array} \right),
\eeq		
it is the $\Phi_{00}$ fields that do not receive masses at one-loop. This is due to the fact that the $\phi$ and $\tilde{\phi}$ fields to which they couple in the superpotential do not get SUSY breaking masses from the non-zero $F$-terms of $\Phi$. The two-loop contribution to their mass near the origin is:
\beq
m_{\Phi_{00}}^2 = -h^2\mu^2 \left( \frac{\alpha_h}{4\pi} \right)^2 N(N_c -N_{f0})\left( 1 + \frac{\pi^2}{6} -\log{4} \right) {\rm Tr}(\Phi_{00}^{\dagger}\Phi_{00}).
\eeq    	
Furthermore, for $\langle \Phi_{00} \rangle >>  \mu$ one finds
\beq
V = -(N_{f1}-N)N h^2\mu^2 \left( \frac{\alpha_h}{4\pi} \right)^2 {\rm Tr} \left( \log {\frac{\Phi_{00}^{\dagger}\Phi_{00}}{\mu^2}} \right)^2 + h^2\epsilon^2\mu^2 {\rm Tr}(\Phi_{00}^{\dagger}\Phi_{00}).
\eeq
So,  $\langle \Phi_{00} \rangle \sim \frac{\mu}{\epsilon}$ and $R$-symmetry is spontaneously broken, since $R(\Phi_{00}) = 1$.  Since this takes place in the magnetic theory, the consistency condition that $\langle \Phi_{00} \rangle << \Lambda$, the condition that higher order terms in the Kahler potential are negligible, and the constraint that the origin is destabilized, demand that $\frac{\mu}{\Lambda} << \epsilon << \frac{\alpha_h}{4 \pi}$.

If this model is used for direct gauge mediation, a subset of the unbroken flavor symmetry can be gauged. However, gaugino masses again vanish to leading order and are generated at higher orders in $\frac{F}{M^2}$.

\subsubsection{Explicit $R$-symmetry breaking}

A different approach to generating gaugino masses is to break the $R$-symmetry explicitly. Many models accomplish this by introducing non-renormalizable operators to the electric superpotential:
\beq \label{explicitR}
\delta W = \frac{{\rm Tr}[(\tilde{Q}Q)^2]}{M_*} + \frac{({\rm Tr}[\tilde{Q}Q])^2}{M_*} + ...
\eeq
where the ``..." refer to higher order terms. The effect of single trace terms on the vacuum structure of SQCD were analyzed in \cite{Giveon:2007ef}.The terms shown become,
\beq \label{magexpR}
\delta W = \Lambda \epsilon_1 {\rm Tr}[\Phi^2] + \Lambda \epsilon_2 ({\rm Tr}[\Phi])^2
\eeq
in the magnetic theory, and are naturally small ($\epsilon_{1,2} \sim \frac{\Lambda}{M_*}$). These terms, in combination with the linear term arising from the ISS mass term, explicitly break the $R$-symmetry and generally lead to SUSY preserving vacua in the magnetic effective theory. Since the coefficients tend to be small, however, the SUSY breaking minima can be sufficiently long lived.

The phenomenology of these deformations was analyzed in \cite{Essig:2008kz}. There it was shown that the double trace term with coefficient $\epsilon_2$ is needed to give mass to adjoint fermions, when one attempts to use this theory as a model of direct gauge mediation. In that case, due to the explicit $R$-symmetry breaking, gaugino masses are generated at leading order. SUSY vacua are also introduced, but they can be sufficiently far off in field space that the SUSY breaking vacua are long lived.

In \cite{Kitano:2006xg} a particular choice of a single trace operator was analyzed:
\beq
\delta W = \frac{{\rm Tr}(\tilde{Q}^{Ia}Q_{a}^J)^2}{M_*}.
\eeq
where $I,J = 1,..,(N_f-N_c)$ and $a,...,N_c$ (see also \cite{Xu:2007az,Zur:2008zg,Kitano:2010fa}). Assuming that the theory preserves an $SU(N_f -N_c) \times SU(N_c)$, the general ISS mass term becomes:  $W = \bar{m}(\tilde{Q}^IQ_I) +\bar{\mu}(\tilde{Q}^aQ_a)$. This interaction introduces new vacua with lower energy than the ISS vacuum, but for $\bar{\mu} \ll \bar{m}$ the original ISS vacuum can be made sufficiently long lived. As a result of the explicit $R$-symmetry breaking, gaugino masses are generated at leading order, but are proportional to $\epsilon \sim \frac{\Lambda}{M_*}$.

Similarly, in \cite{Giveon:2009yu} a partial trace term was considered. In particular, the $N_{f0}$ massless flavors of \cite{Giveon:2008ne} (\eq{Giveon}) were given a small mass $W = m_0 \delta_{ij}\tilde{Q}^iQ^j$ ($m_0 < m_1$) and $\epsilon_1 = 0$ in \eq{magexpR}. In this case,
\beq
W = \mu_1^2{\rm Tr}(\Phi_{11})+ \mu_0^2{\rm Tr}(\Phi_{00}) + h\tilde{\phi}_i\Phi^{ij}\phi_j + \frac{h}{2}\epsilon \mu_0 {\rm Tr}(\Phi_{00}^2).
\eeq
The theory can be analyzed in the vacuum where the vev of the $N_{f0}$ magnetic quarks is zero. Together the mass term and the linear term explicitly break the $R$-symmetry and lead to nonzero vev for $\Phi_{00}$, $\mu_0 \langle \Phi_{00} \rangle \mu_1$. This region is locally stable if $\sqrt{k(N-k)}\frac{\alpha}{4\pi}\frac{\mu_0}{\mu_1} < \epsilon < k\frac{\alpha}{4\pi}\frac{\mu_0}{\mu_1}$. This condition arises from the balance between one-loop potentials for the
dual quark fields that couple to $\Phi_{00}$ (which generate a $\log{|Z|^2}$ potential) and the dual quark fields that couple to $\Phi_{01}$ and $\Phi_{10}$, which generate a quadratic potential. Thus one finds that there is an additional SUSY breaking state which is metastable with respect to the ISS vacuum of the theory (the one where all $N$ dual quarks have vev's). Now, with a locally stable vacuum, one can gauge a global flavor symmetry and derive that gaugino masses are generated at leading order. This is one of the few models where direct mediation can generate gaugino masses at leading order. However, the large number of extra fields in this type of scenario generically eliminates the possibility of perturbative gauge coupling unification. In \cite{Green:2010ww} an alternative approach achieved scalar masses screening in addition to gaugino mass screening in order to keep the gaugino/scalar mass ratio of order one.

In \cite{Barnard:2009ir} similar types of vacua were examined within in the context of the ``baryon deformed models." In this case the explicit $R$-symmetry breaking operators in the electric theory are only single trace operators of the form
\beq
\delta W = \frac{{\rm Tr}(\tilde{Q}Q)^2}{M_*} + \frac{{\rm Tr}(\tilde{Q}Q)^3}{M^3_*}.
\eeq
If one takes the $\frac{\Lambda}{M_*} \sim 10^{-1}$, then these theories allow for a locally stable vacuum, where the magnetic quarks have no vev. This is one of the small class of theories where gauginos have mass at leading order. This model in particular results in substantial values for the gravitino mass.

Finally, in \cite{Haba:2007rj}, a higher dimensional baryon deformation was considered, which manifests itself in the magnetic theory as
\beq
\delta W = \frac{{\rm Tr}[(\tilde{q}q)^2]}{\Lambda_{UV}}.
\eeq
This term explicitly breaks $R$-symmetry, introduces SUSY vacua at large fields vev's, and generates gaugino masses at subleading order in the $\frac{F}{M^2}$ expansion.

So far, we have discussed deformations in the context of the electric theory with $SU(N)$ gauge symmetry. In \cite{Koschade:2009qu}, it was shown that similar deformations can be successfully used for model building within the framework of $SO(N)$ theories.

\subsubsection{The ISS mass term}

The ISS analysis is only valid when the mass term in the electric theory and the dynamical scale obey the relation: $m_Q \ll \Lambda$; calculations are only reliable in this limit. In a theory where all mass scales are truly dynamically generated, the origin of this mass scale requires an explanation. Models address this with a second dynamical scale generated from dimensional transmutation of an auxiliary sector of strong dynamics $\Lambda_{aux}$. In practice this approach is very similar to retro-fitting.

In order to address the smallness of the quark mass in the electric theory of \cite{Csaki:2006wi}, an $N_f = N_c = 2$ sector was added to the theory which dynamically generated a scale from its quantum deformed moduli space\cite{Seiberg:1994bz,Seiberg:1994pq}. One can then re-write the original ISS mass term, $m {\rm Tr}(\tilde{Q}Q)$, as: $\frac{\bar{p}p}{\Lambda_{UV}^3}{\rm Tr}(\tilde{Q}Q)$, where $p(\bar{p})$ represents a baryon of the $SU(2)_{aux}$ theory. On the baryon branch, one finds a SUSY breaking scale of $F = \frac{\Lambda \Lambda_{aux}^4}{\Lambda^3_{UV}}$, where $\Lambda$ is the dynamical scale of the ISS theory. A similar coupling can be used to generate the $\mu$ term. Taking $W =  \frac{\bar{p}p}{\Lambda_{UV}^3}H_UH_D$, one finds $\mu = \frac{\Lambda_{aux}^4}{\Lambda_{UV}^3}$.

A different approach was used in \cite{Brummer:2007ns,Essig:2007xk}, where an auxiliary sector of $SU(N_c')$ with $N_f'<N_c'$ flavors was added to an $SU(N_c)$
 theory with $ N_c+1 < N_f < \frac{3}{2}N_c$ flavors, with a singlet ($\phi$) field coupling the two theories. In this case the interactions are \footnote{Reference \cite{Brummer:2007ns} included a cubic interaction}:
\beq
W = \phi {\rm Tr} (\tilde{Q}Q) + \phi {\rm Tr} (\tilde{P}P).
\eeq
Below the scale of the strong dynamics of both theories, the $SU(N_c')$ theory generates a potential for the singlet from the ADS superpotential and the $SU(N_c)$ theory is described in terms of the magnetic, free, theory of ISS but with the mass term replaced with the vev of the singlet. Expanding around the ISS vacuum, one finds that the singlet vev is stable with a dynamically small vev. Including the contribution from the Coleman-Weinberg potential then stabilizes the would-be flat direction of the ISS meson to a small but non-zero value. This vacuum breaks the $R$-symmetry spontaneously and can be used for gauge mediation. In \cite{Brummer:2010zx} an alternate usage of and extra gauge group was used to dynamically generate a small ISS mass term.

\subsubsection{Landau Poles}

In all of the models where a flavor subgroup of the ISS sector is gauged under the Standard Model gauge group, there will be matter charged under the standard model gauge group in addition to the MSSM fields. If this extra matter is sufficiently light, the SM gauge couplings will hit Landau poles in the UV well below the GUT scale. One possibility is to allow some of the extra matter fields to pair up with spectator fields and forcibly remove them from the low-energy spectrum \cite{Franco:2009wf,Craig:2009hf}. Gauging a ``chiral" subgroup of the flavour group can also reduce the number of fields contributing to the Standard Model beta function \cite{Behbahani:2010wh}.

It is also plausible that the Standard Model itself has a dual description in terms of another calculable theory above the scale of the Landau Pole. This is the attitude taken in \cite{Abel:2008tx,Abel:2009bj}, where it was shown that an $SU(11)\times Sp(1)^3$ theory has a dual description as the MSSM with messengers.  This was proposed as a plausible picture of an ISS Sp-type theory above the Landau Pole; however a fully calculable model of this type remains to be found.

In the class of models where the ISS mass term and meson deformations do not respect a full $SU(N_f)$ flavor symmetry, one can include a separation of scales between the mass scale of SUSY breaking and the mass scale of some of the extra matter charged under the Standard Model. In particular, the pseudo-moduli of the meson can contribute significantly to the RGE of the SM gauge couplings. However, in models where the flavor symmetry is broken, $SU(N_f) \rightarrow SU(N_f-N_C)\times SU(N_c)$ (\cite{Kitano:2006xg} for example), the lighter meson pseudo-moduli are not charged under the $SU(N_f-N_c)$ gauge group left unbroken by the dual quark vev's. Embedding the SM in this $SU(N_f-N_c)$ flavor symmetry, can reduce the number of light matter fields charged under the Standard Model.

\subsubsection{Modular Gauge Mediation}

A modular approach to gauge mediation can avoid many of the problems encountered when attempting to gauge the flavor symmetry of the ISS model. In \cite{Murayama:2006yf,Aharony:2006my,Murayama:2007fe} a very simple class of models of the form
\beq \label{mnas}
W = m_{ij}\tilde{Q}^iQ^j + \lambda_{ij}\frac{\tilde{Q}^iQ^j\tilde{f}f}{M_{pl}} + M\tilde{f}f
\eeq
were shown to be complete versions of gauge mediation. Here $f$($\tilde{f}$) are vector-like pairs of fundamentals under the Standard Model gauge group and play the role of the messengers. $m_{ij}$ is the just the ISS mass term. If the $Q$($\tilde{Q}$) are $N_f$ matter fields of an $SU(N_c)$ ($N_c < N_f < \frac{3}{2}N_c$) with dynamical scale $\Lambda$, then the effective theory below $\Lambda$ is:
\beq
W = FX + \lambda' X\tilde{f}f + M\tilde{f}f,
\eeq
(where we have taken, for simplicity, $m_{ij}= m \delta_{ij}$, $\lambda_{ij} = \lambda \delta_{ij}$, and ${\rm Tr}(\Phi) = X$) with $F = \Lambda m$ and $\lambda' = \frac{\Lambda}{M_{pl}}$.
This model can be viewed as one that explicitly breaks $R$-symmetry, but the smallness of the coupling $\lambda'$ implies that the SUSY vacuum with $\langle \tilde{f}f \rangle \not= 0 $ is far from the metastable vacuum.

The two mass scales, $m$ and $M$, can have a common origin from gaugino condensation. The authors of \cite{Aharony:2006my} propose to replace:
\beq
m {\rm Tr}(\tilde{Q}Q) + M\tilde{f}f \rightarrow \frac{W^{\alpha}W_{\alpha}}{M^2_{pl}}( {\rm Tr}(\tilde{Q}Q) + \tilde{f}f ),
\eeq
where $W^{\alpha}W_{\alpha}$ is the gauge kinetic term from a pure SUSY Yang-Mills sector. After the gauginos condense one has that $m \sim M \sim \frac{\Lambda^3}{M_{pl}^2}$. Upon requiring the SUSY scale to be $1~ {\rm TeV}$ due to typical gauge mediated loops of messenger fields, one finds that the simple dynamical model above yields a fundamental SUSY breaking scale of $\sqrt{F} \sim 10^3~{\rm TeV}$.

An alternative to the above is to utilize any model which incorporates spontaneous $R$-symmetry breaking and couple messengers with the non-renormalizable operator in  \eq{mnas}. This is also sufficient to generate a theory of modular gauge mediation.

\subsection{Retrofitting the O'Raifeartaigh Models}
\label{retrofitting}

In this section, we will describe a strategy for building models with metastable, dynamical supersymmetry breaking, which
is in many ways simpler than the ISS construction we have discussed up to now.
The basic ingredient, as we will see, are models which dynamically break a discrete $R$ symmetry, without breaking
supersymmetry.  The prototype for such theories are pure $SU(N)$ gauge theories, in which gaugino condensation
breaks a $Z_N$ symmetry.  We have seen that one can readily generalize gaugino condensation
to theories which include order parameters of dimension one.  In this section, we will see that using these ingredients,
one can {\it very} easily build models
in which all dimensionful parameters arise through dimensional transmutation, including the $\mu$ term of the MSSM,
and possible parameters of the NMSSM.  We will also note a special feature of $\langle W \rangle$ in this framework.
Since
$W$ transforms under any $R$ symmetry, the expectation value of the superpotential
is itself an order parameter for $R$ breaking.  In the context of supergravity
theories, this is particularly important.  In the retrofitted models, the relations among scales are of the correct
order of magnitude to account for the smallness of the cosmological constant; this is not true of many other schemes for supersymmetry
breaking, where additional scales must be introduced by hand.

\subsubsection{Gauge Mediation/Retrofitting}
\label{generalized}

Given our models of gaugino condensation, it is a simple matter to generate the various dimensionful couplings of O'Raifeartaigh models dynamically.
In the model of \eq{simplestor}, for example, we can make the replacements:
 \beq
X (A^2 - \mu^2) + m AY \rightarrow{X W_\alpha^2 \over M_p}+ \gamma S A Y.
\label{retrofittedmodel}
\eeq
Note that $\langle W \rangle  \approx  \Lambda^3$, $\langle S \rangle \sim \Lambda$, and
$m^2 \gg f$.
As in our earlier perturbed O'Raifeartaigh models, this model has supersymmetric
minima.  If we suppose that this structure is enforced by a discrete symmetry,
the addition of higher order terms will allow such vacua.  Even if not, however,
for large $X$ the superpotential behaves as $e^{-X/b_0}$, and tends to zero at $\infty$.
On the other hand, near the origin of field space, the Coleman-Weinberg calculation goes through as before,
and the potential has a local, supersymmetry-breaking minimum.  Because the supersymmetric vacuum
is far away, this metastable state is stable.

This model has other interesting features.  It has a, presumably approximate, continuous $R$ symmetry at low
energies.
If we wish to account for this as a consequence of a discrete $R$ symmetry, at a microscopic level,
the field $X$ must be neutral.  So the exact, discrete symmetry is not a subgroup of the (approximate) continuous
symmetry.  A candidate symmetry might be a $Z_N$.  Defining $\alpha = e^{2 \pi i \over N}$, we might assign
transformation laws:
\beq
X \rightarrow X;~~~~  A \rightarrow \alpha A;~~~~ Y \rightarrow \alpha Y.
\eeq
As before, we require, say, a $Z_2$ under which $A$ and $Y$ are odd to completely
account for this structure.

\subsubsection{Gauge Mediation and the Cosmological Constant}

A traditional objection to gauge mediated models\footnote{T. Banks, unpublished.}
is that the smallness of the c.c. requires a large constant in $W$, unrelated to anything else.
But we have just seen that in retrofitted models, one naturally expects $\langle W \rangle \approx F M_p^2$, i.e. of the correct order of magnitude to (almost) cancel
the SUSY-breaking contributions to the c.c.  This makes retrofitting, or something like it,
almost inevitable in gauge mediation.

\subsubsection{Improving the Simplest Model}

The model of \eq{retrofittedmodel} suffers from several shortcomings:
\begin{enumerate}
\item  It still possesses a parameter with dimensions of mass, $m$.
If we try to retrofit this using gaugino condensation, we will have $m \sim { \Lambda^3 \over M_p^2}$,
which is problematic for model building (a first effort to circumvent this difficulty appeared in \cite{Dine:2006xt}).
\item  The model possesses an approximate, continuous $R$ symmetry, which is unbroken.
\item  If one attempts to develop this model into a full theory of gauge mediation, one needs to account
for other dimension one mass terms, such as the $\mu$ term.
\end{enumerate}

The first problem can be resolved by coupling singlets, of the sort discussed in section \ref{generalized}.   Depending on the choice of scales,
one can contemplate, for example, replacing $mAY$ with
\beq
\gamma S A Y;~~~~~ {S^2 \over M} AY.
\eeq
Here, for example, $\gamma$ is a dimensionless number.  The second problem can be solved by retrofitting a model like that of Shih, \eq{shihmodel}
The required structure can be accounted for by discrete symmetries.
Allowing the $X$ field of \eq{shihmodel} to couple to messengers, one can readily write down models of gauge mediation.

Another longstanding issue in gauge mediation is the $\mu$ problem.  In gravity mediation, the fact that $\mu$ is small, of order the weak scale, can readily
be understood.  Order $M_p$ or $M_{gut}$ contributions can be suppressed by symmetries; in string theory, one often finds analogous masses simply
vanish at tree level, and there are no radiative corrections
due to non-renormalization theorems.  The needed $\mu$ can be generated by effects connected with supersymmetry breaking\cite{Giudice:1988yz}.  In these cases, $B_\mu$, the soft
breaking Higgs mass term in the MSSM lagrangian, is of order $\mu^2$.  In gauge mediation, however, if $\mu$ is generated by loop effects
similar to those which generate soft breaking terms, $B_\mu$ is parameterically too large.  E.g. if $\mu$ is a two loop effect, so is $B_\mu$, and so
\beq
B_\mu \gg \mu^2.
\eeq

In retrofitted models, particularly in the presence of dimension one order parameters,
 the $\mu$-term, arises readily from dynamical breaking of the discrete $R$ symmetry\cite{Yanagida:1997yf,Dine:2009swa}.  For example,
\beq
W_{\mu} =
 \gamma{S^2 \over M_p} H_U H_D\eeq
can readily yield a $\mu$ term of a suitable order of magnitude.  Because the $F$ term of $S$ is very small, this generates $\mu$ without $B_\mu$.
$B_\mu$ is then generated by one loop running, and is in fact small, leading to large values of $\tan \beta$.

\section{Conclusions:  Supersymmetry, Supersymmetry Breaking, and Nature}

As this is being written, the LHC program is just beginning.  One is likely to know, soon, whether superparticles exist with masses less than about $700$ GeV or so.
Supersymmetry remains perhaps the best motivated proposal for new physics which might account for the hierarchy, and which should appear, if not in the first round
of LHC experiments, not too long thereafter.  Critical, however, to the plausibility of supersymmetry as a model for the physics which underlies electroweak
symmetry breaking is the possibility of dynamical breaking.  A decade ago, models of dynamical supersymmetry
breaking were, at best, rather unsightly.  We hope to have convinced the reader that, with
the shifted focus to metastable supersymmetry breaking, this situation has drastically changed, and it is possible to write
quite compelling models.  It remains to be seen if nature has the good taste to take advantage of one of these structures.

\noindent
{\bf  Acknowledgements}
This work
was supported in part by the U.S.~Department of Energy.  M.~Dine
thanks Stanford University and the Stanford Institute for Theoretical
Physics for a visiting faculty appointment while some of this work was
performed.  J. Mason was supported by NSF grant PHY-0855591, The Harvard Center for the Fundamnetal Laws of Nature, and would like to thank The Aspen Center for Physics for their hospitality while while working on this project. We thank our many collaborators and friends for discussions of
issues relevant here:  Tom Banks, Guido Festuccia, Jonathan Feng, John Kehayias, Zohar Komargodski, Tongyan Lin, David Poland, Aqil Sajjad, Yuri Shirman, Nathan Seiberg, Eva Silverstein, Zheng Sun and others too numerous to mention.  We particularly thank Sebastian Grab and Zohar Komargodski for their careful reading of the manuscript.

\bibliography{susyrefs1213}{}
\bibliographystyle{utphys}
\end{document}